\documentclass[aps,superscriptaddress,twocolumn,pre,longbibliography,floatfix]{revtex4-1}
\usepackage{amsmath,amsthm,amsfonts,amssymb,bm,graphicx,color,mathpazo,times, mathtools, dsfont}
\usepackage[colorlinks={true}, citecolor={blue}, filecolor={blue}, linkcolor={blue}, urlcolor={blue}]{hyperref}
\usepackage[caption=false]{subfig}
\usepackage{soul}
\usepackage{braket}

\newcommand{\abs}[1]{\left\vert#1\right\vert}

\allowdisplaybreaks

\begin{document}

\preprint{APS/123-QED}

\title{Two-body quantum absorption refrigerators with optomechanical-like interactions}

\author{M.\ Tahir Naseem}
\affiliation{Department of Physics, Ko\c{c} University, Sar{\i}yer, \.Istanbul, 34450, Turkey}
\author{Avijit Misra}
\affiliation{Department of Chemical and Biological Physics, Weizmann Institute of Science, 
		Rehovot 7610001, Israel}
\author{\"{O}zg\"{u}r E.\ M\"{u}stecapl{\i}o\u{g}lu}
\email[Electronic address:\ ]{omustecap@ku.edu.tr}
\affiliation{Department of Physics, Ko\c{c} University, Sar{\i}yer, \.Istanbul, 34450, Turkey}
\date{\today}

\begin{abstract}
Quantum absorption refrigerator (QAR) autonomously extracts heat from a cold bath and dumps into a hot bath by exploiting the input heat from a higher temperature reservoir. QARs typically require three-body interactions.  We propose and examine a two-body QAR model based upon optomechanical-like coupling in the working medium composed of either two two-level systems or two harmonic oscillators or one two-level atom and a harmonic oscillator. In the ideal case without internal dissipation, within the experimentally realizable parameters, our model can attain the coefficient of performance that is arbitrarily close to the Carnot bound. We study the efficiency at maximum power, a  bound for practical purposes, and show that by using suitable reservoir engineering and exploiting the nonlinear optomechanical-like coupling, one can achieve  efficiency at maximum power close to the Carnot bound, though the power gradually approaches zero as the efficiency approaches the Carnot bound. Moreover, we discuss the impact of non-classical correlations and the size of Hilbert space on the cooling power. Finally, we consider a more realistic version of our model in which we consider heat leaks that makes QAR non-ideal and prevent it to achieve the Carnot efficiency. 
\end{abstract}

\maketitle

\section{\label{sec:Intro}Introduction}
Thermal machines with quantum working substances, such as quantum heat engines~\cite{doi:10.1146/annurev-physchem-040513-103724, Rossnagel325, PhysRevE.96.062120} and refrigerators~\cite{PhysRevB.94.184503, PhysRevApplied.11.054034}, have been attracted much attention recently. A particularly promising class of quantum thermal machines are those that do not require any external work. For compact and energetically cheap implementations of quantum technologies, autonomous quantum cooling schemes are highly desirable. For that aim, so-called quantum absorption refrigerators (QARs)~\cite{doi:10.1080/00107514.2019.1631555} have been proposed~\cite{PhysRevE.64.056130, PhysRevLett.105.130401, PhysRevLett.108.070604, PhysRevE.85.061126, PhysRevE.87.042131,PhysRevLett.110.256801,PhysRevE.89.042128,PhysRevA.91.012117,PhysRevE.92.012136,PhysRevE.94.032120,PhysRevB.94.235420,PhysRevE.93.022134,PhysRevE.95.062131,PhysRevB.98.045433,PhysRevB.98.081404,PhysRevE.97.052145,PhysRevE.98.012117,PhysRevE.98.012131,Mitchison_2016,Chen_2012, PhysRevE.96.052126, PhysRevE.97.062116, Das_2019, Latune2019, PhysRevE.101.012109} and recently a three-ion QAR has been experimentally demonstrated~\cite{Maslennikov2019}. 
While typical QARs rely upon three-body interaction~\cite{PhysRevLett.108.070604, Chen_2012}, smaller QARs, such as two-atoms~\cite{PhysRevLett.105.130401, PhysRevE.85.061126} or two-resonators~\cite{PhysRevE.85.061126}, have also been proposed.

The ideal QAR based on two-body interaction can saturate the Carnot bound, at the cost of vanishing cooling power~\cite{doi:10.1080/00107514.2019.1631555}. A more practical bound for QAR then can be efficiency at maximum cooling power $\epsilon_{*}$~\cite{Correa2014}. Surprisingly, such a bound has been established in~\cite{Correa2014} for the QARs that can be broken up in three-level masers. 
Two-body QAR~\cite{PhysRevE.85.061126} has been shown to obey this bound, which is much smaller than the Carnot efficiency. Here we consider two-body QAR with directly interacting subsystems of the working medium. We show that by using reservoir engineering, our model can surpass this bound. In the strong coupling regime, the efficiency at maximum power even can saturate the Carnot efficiency (the power gradually goes to zero at the Carnot point). In addition to discrete and continuous variable (CV) models, we investigate a hybrid system in which a two-level atom is coupled with a single resonator. The coupling then becomes identical to the single-mode case of electron-phonon~\cite{frochlich_interaction_1952,holstein_studiesI_1959} or spin-boson interaction~\cite{weiss_quantumDissSys_4ed2012}. In our model, we observe that larger Hilbert space sizes enhance the cooling power~\cite{PhysRevE.89.032115}, and quantum correlations play no role in the effecient performance of the refrigerator~\cite{PhysRevE.87.042131}.
Finally, we consider a more realistic case, in which our model suffers from heat leaks, consequently becomes non-ideal~\cite{PhysRevE.64.056130}.

To give analytical and physical explanations of our results, we use the atomic analog of the optomechanical model. Such two-atom QARs can be of interest per se and can be realized by quantum optical methods~\cite{PhysRevE.99.042121}. Using the atomic model, we explain the cooling mechanism by the energy cycles that consist of thermally induced transitions between the dressed states of the working medium. We identify similar transition structures within CV models to justify the emergence of the same cooling mechanism in them. In suitable parameters regime, according to the second law of thermodynamics, the energy cycles that remove the heat from the cold bath become more favorable than the reciprocal process. We use a thermodynamically consistent master equation that has been recently derived for optomechanical systems~\cite{PhysRevA.98.052123}. We verify that proposed devices operate within the Carnot bound but can get arbitrarily close to Carnot efficiency, where unfortunately but necessarily the cooling power vanishes.

The rest of the paper is organized as follows. We describe the three models of QARs that we examine in Sec.~\ref{sec:jmodel}. They are all based upon the standard nonlinear optomechanical model, which is the pure CV case. The other two cases are for its hybrid and atomic analogs. The mechanism of the cooling is explained for the three working mediums in Sec.~\ref{sec:Mechanism}, where main analytical results are based upon exactly solvable atomic analog. Detailed analysis of thermally induced transitions among the dressed states is given and compared for justification of the common physics behind QAR for the three models.  We discuss non-ideal model of our QAR in Sec.~\ref{sec:heatleak} and conclude in Sec.~\ref{sec:conclusions}.  

\section{Quantum absorption refrigerator model}\label{sec:jmodel}
	\begin{figure}[t]
		\centering
		\includegraphics[width=0.5\textwidth]{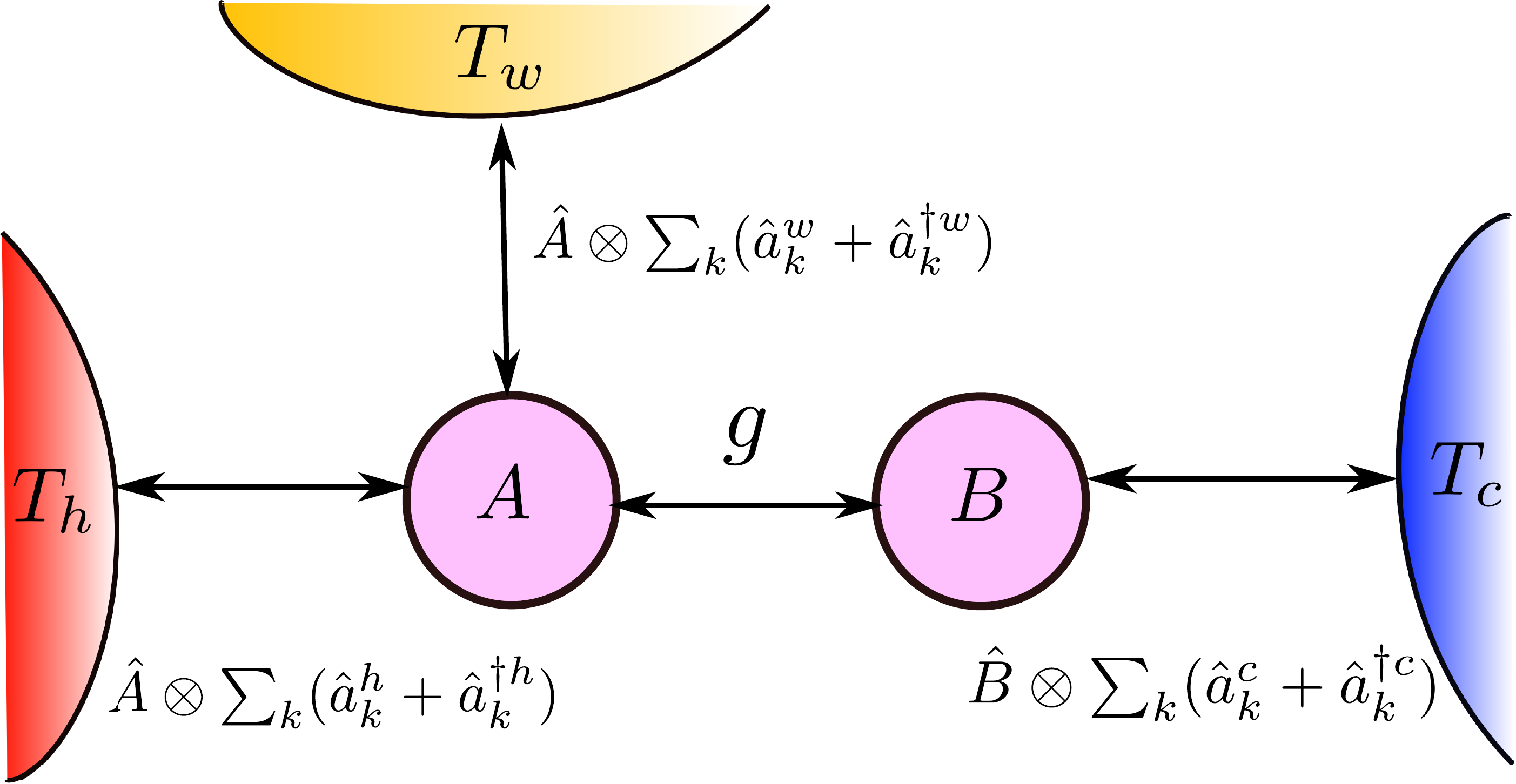}
		\caption{(Color online) Schematic illustration of a quantum absorption refrigerator in which the working medium is composed of two coupled qubits or two coupled harmonic oscillators or a single qubit (A) and a harmonic oscillator (B). The frequency of subsystem A (B) is $\omega_h$ ($\omega_{c}$),  and the coupling strength between them is $g$. The subsystem A is coupled with two thermal baths, hot bath and work reservoir namely, at temperatures $T_{h}$ and $T_{w}$,  and subsystem B is coupled with the cold bath at temperature  $T_{c}$. We also have $T_{w}>T_{h}>T_{c}$.}
		\label{fig:jmodel}
	\end{figure} 

We propose a model of QAR in which the working medium is composed of two coupled systems (A) and (B), as shown in Fig.~\ref{fig:jmodel}. We consider three different cases viz. (i) A and B both are two-level systems (TLS) (ii) A is a TLS and B is a harmonic oscillator and (iii) Both A and B are harmonic oscillators. The subsystem A is coupled with a hot bath at temperature $T_{h}$, besides, it is also coupled with a thermal bath that plays the role of work reservoir with a temperature of $T_{w}$. 
The subsystem B is coupled with a cold bath at temperature $T_{c}$. We assume that all three thermal baths are independent with temperatures $T_{w}>T_{h}>T_{c}$. 

The Hamiltonian of the working medium (A + B) of our QAR is given by~\cite{RevModPhys.86.1391}
\begin{equation}\label{eq:Hamsys}
\hat{H}_{S}= \omega_{h}\hat{a}^{\dagger}\hat{a} + \omega_{c}\hat{b}^{\dagger}\hat{b} + g \hat{a}^{\dagger}\hat{a} (\hat{b}+\hat{b}^{\dagger}),
\end{equation} 
where $g$ is the coupling strength, $\omega_{h}$ and $\omega_{c}$ are the frequencies of the A and B subsystems, respectively. The bosonic annihilation (creation) operator for the subsystem A is represented by $\hat{a}$ ($\hat{a}^{\dagger}$), and those of subsystem B with $\hat{b}$ ($\hat{b}^{\dagger}$). For the case of coupled two two-level system (TLS), $\hat{a},\hat{a}^{\dagger}$ and $\hat{b},\hat{b}^{\dagger}$ are replaced Pauli operators $\hat{\sigma}^{\pm}_{h, c}$, and the last term becomes atomic analog of optomechanical  coupling, $g\hat{\sigma}^{z}_{h}\hat{\sigma}^{x}_{c}$, which can be realized by quantum optical methods~\cite{PhysRevE.99.042121}. For case of two-level-oscillator system (TLOS), we replace $\hat{a},\hat{a}^\dag$ by the Pauli operators $\hat{\sigma}^{\pm}_{h}$, yielding hybrid optomechanical-like model $g\hat{\sigma}_h^z(\hat{b}+\hat{b}^{\dagger})$, which is identical with the single-mode case of electron-phonon~\cite{frochlich_interaction_1952,holstein_studiesI_1959} or spin-boson coupling~\cite{weiss_quantumDissSys_4ed2012}. 

The Hamiltonian of the independent thermal baths is given by
\begin{equation}\label{Eq:HamBaths}
\hat{H}_{B\alpha}= \sum_{\mu,\alpha}\omega_{\mu}\hat{a}^{\dagger}_{\mu,\alpha}\hat{a}_{\mu,\alpha},
\end{equation}  
where sum is taken over infinite number of bath modes, indexed by $\mu$, and $\alpha = w, h, c,$  representing the work-like reservoir, hot and cold thermal baths, respectively.  The interaction of system (A+B) with the baths are defined as
\begin{multline}
\hat{H}_{SB}=\sum_\mu g_{k, \mu}(\hat{a}+\hat{a}^{\dagger})\otimes(\hat{a}_{k,\mu}+\hat{a}^{\dagger}_{k,\mu}) \\ + \sum_\mu g_{c,\mu}(\hat{b}+\hat{b}^{\dagger})\otimes(\hat{a}_{c,\mu}+\hat{a}^{\dagger}_{c,\mu}), 
\end{multline} 
where $k = w, h$. In case of TLS the term $(\hat{a}+\hat{a}^{\dagger})$ is replaced by $\sigma^x$.

In order to derive the master equation, first we diagnolize the Hamiltonian given in Eq.~(\ref{eq:Hamsys}) using the transformation
\begin{equation}\label{eq:unitary}
\hat{U} = e^{-\beta \hat{a}^{\dagger}\hat{a}(\hat{b}^{\dagger}-\hat{b})},
\end{equation}
where $\beta = g/\omega_{c}$ for both hybrid and pure optomechanical systems (OMS). It represents $\beta = \theta/2$ for TLS, where the angle is defined by
$\sin\theta := 2g/\tilde{\omega_{c}}$, $\cos\theta := \omega_c/\tilde{\omega_{c}}$ and $\tilde{\omega_{c}} = \sqrt{\omega^2_{c}+4g^2}$.
The master equation in the interaction picture is given by
\begin{eqnarray}\label{eq:master}
&\dot{\tilde{\rho}}& = \hat{\mathcal{L}}_{w} +\hat{\mathcal{L}}_{h} +\hat{ \mathcal{L}}_{c},
\end{eqnarray}
where $\hat{\mathcal{L}}_{\alpha}$ are Liouville superoperators that describe the energy exchange of the working medium with the baths, which are of the form ~\cite{PhysRevE.99.042121, PhysRevA.98.052123, PhysRevE.87.012140} 
\begin{eqnarray}
	\label{eq:L_L}
\hat{ \mathcal{L}}_{k} &=& G_{k}(\omega_{h})c^{2}\hat{\mathcal{D}}[\tilde{a}]
	+ G_{k}(-\omega_{h})c^{2}\hat{\mathcal{D}}[\tilde{a}^{\dagger}] \\ \nonumber&+& G_{k}(\omega_{w})s^{2}\hat{\mathcal{D}}[\tilde{a}\tilde{b}^{\dagger}]
	+ G_{k}(-\omega_{w})s^{2}\hat{\mathcal{D}}[\tilde{a}^{\dagger}\tilde{b}]
	\\ \nonumber &+& G_{k}(\omega_{+})s^{2}\hat{\mathcal{D}}[\tilde{a}\tilde{b}]
	+ G_{k}(-\omega_{+})s^{2}\hat{\mathcal{D}}[\tilde{a}^{\dagger}\tilde{b}^{\dagger}],
	\\
	\hat{ \mathcal{L}}_{c}&=& G_{c}(\tilde{\omega}_{c})c^{2}\hat{\mathcal{D}}[\tilde{b}]
	+ G_{c}(-\tilde{\omega}_{c})c^{2}\hat{\mathcal{D}}[\tilde{b}^{\dagger}], \label{eq:L_M}
\end{eqnarray}
where $\omega_{+}=\omega_{h}+\tilde{\omega}_{c}$, and $\omega_{w}=\omega_{h}-\tilde{\omega}_{c}$. For TLS, we write $c^2=\cos^2\theta$, and $s^2=\sin^2\theta$; while for both hybrid and pure optomechanical systems, we like to emphasize, the notations represent $\tilde{\omega}_{c}=\omega_{c}$, $c^2 =1$, and $s^2 =(g/\tilde{\omega}_{c})^2$. We ignore the pure decoherence (dephasing) dissipators, because they do not change the diagonal elements of the density matrix in the energy eigenbasis and therefore, do not contribute to steady-state heat flow.
We present the explicit expressions for the transformed operators $\tilde{a}$ and $\tilde{b}$ with eigenvalues and eigenvectors for the dressed Hamiltonians in the Appendix~\ref{AppendixA}. 
The bath spectral response functions are given by
\begin{eqnarray}\label{eq:SRF}
G_{\alpha}(\omega)=
\begin{cases}
\gamma_{\alpha}(\omega)[1 + \bar{n}_{\alpha}(\omega)] & \omega > 0, \\
\gamma_{\alpha}(\abs{\omega})\bar{n}_{\alpha}(\abs{\omega}) & \omega < 0,
\end{cases}
\end{eqnarray} 
and $\bar{n}_{\alpha}(\omega) := 1/(\exp{\omega/T_{\alpha}} - 1)$ is the average excitation number of the baths, and the energy damping rate~\cite{RevModPhys.59.1}
\begin{eqnarray}\label{eq:BSD}
\gamma_{\alpha}(\omega) &=& 2\pi\sum_{\mu}g^2_{\mu,\alpha}\delta(\omega-\omega_{\mu,\alpha})=2\pi J_{\alpha}(\omega), and \nonumber\\
J_{\alpha}(\omega) &=& \kappa_{\alpha} \frac{\omega^{s}}{\omega_{\text{ct}}^{1-s}} e^{-\omega/\omega_{\text{ct}}}, 
\end{eqnarray}
where, $\omega_{\text{ct}}$ is the cutoff frequency. In this work, we consider the Ohmic ($s=1$) spectral density (OSD) of the baths, unless otherwise mentioned. The Lindblad dissipators in the Eqs.~(\ref{eq:L_L}) and (\ref{eq:L_M}) $\hat{\mathcal{D}}[\hat{o}]$ are defined as
\begin{equation}\label{dissipator}
\hat{\mathcal{D}}[\hat{o}] = \frac{1}{2}(2\hat{o}\hat{\rho}\hat{o}^{\dagger} - \hat{o}^{\dagger}\hat{o}\hat{\rho} + \hat{\rho}\hat{o}^{\dagger}\hat{o}).
\end{equation}
We assume OSD for the cold bath, and spectrally filtered response functions for the work-like and hot reservoirs, which is taken to be~\cite{doi:10.1080/09500349414550381, PhysRevE.90.022102}
\begin{equation}\label{eq:BathFilter}
G^{k}_f = \frac{\kappa_{f}}{\pi}\frac{(\pi G^{k}(\omega))^2}{[\omega-\big(\omega^{k}_{f}+\Delta^{k}(\omega)\big)^2]+(\pi G^{k}(\omega))^2},
\end{equation}
where $G^{k}(\omega)$ is unfiltered coupling spectrum, and
\begin{equation}
\Delta^{k}(\omega) = P \bigg[\int^{\infty}_{o} d\omega^{'} \frac{G^{k}(\omega^{'})}{\omega - \omega^{'}}\bigg],
\end{equation}
where $P$ is the principal value. For the OMS, filtering of the bath spectrum can be achieved using photonic bandgap materials and phononic cavities~\cite{MF}.
 The motivation for using spectrally filtered response functions of hot and work baths are to avoid the unwanted overlap of these baths. This can induce heat-leaks, consequently, the efficiency of the QAR reduces, and the details of which are given in Sec.~\ref{sec:heatleak}.
\section{Ideal QAR}\label{sec:Mechanism}
In this section, we discuss the parameters regime in which our device work as a fridge and underlying mechanism of cooling.
\subsection{The cooling window}
We are interested in the steady-state analysis for our QAR, and the steady-state heat currents from the bath $\alpha$ into the system is given by~\cite{PhysRevE.66.036102} 
\begin{equation}\label{Eq:HeatCurrent}
\mathcal{J}_{\alpha} = \text{Tr}\{(\mathcal{L}_{\alpha}\tilde{\rho})\tilde{H}_{\text{S}}\}.
\end{equation}
The heat flow into the system is taken as positive, and $\tilde{H}_{S}$ is the diagonlized Hamiltonian of the system in terms of transformed operators. 
At steady-state, the first and second law of thermodynamics is respectively written as
\begin{eqnarray}
\sum_{\alpha}\mathcal{J}_{\alpha} &=& 0, \label{eq:firstlaw}\\ 
\sigma = -\sum_{\alpha}\frac{\mathcal{J}_{\alpha}}{T_{\alpha}}&\geq & 0, \label{eq:second}
\end{eqnarray}
where $\sigma$ is the total entropy production 
 of the system and reservoirs at the steady state. 

The figure of merit for the performance of QAR is obtained by comparing the ratio of heat extracted from the cold bath to heat injected into the system from the work-like reservoir. Accordingly, the coefficient of performance (COP) for the absorption refrigerator is defined as $\epsilon = \mathcal{J}_{c}/\mathcal{J}_{w}$. The COP is colloquially referred as `efficiency' of the QAR, although, unlike heat engine it can be greater than one. For  refrigeration, we must have $\mathcal{J}_{w}$, $\mathcal{J}_{c}>0$ and $\mathcal{J}_{h}<0$, then by using the Eq.~(\ref{eq:firstlaw}) to remove the $\mathcal{J}_{h}$ from the Eq.~(\ref{eq:second}), we get
\begin{equation}
\epsilon = \frac{\mathcal{J}_{c}}{\mathcal{J}_{w}}\leq \frac{(T_{w}-T_{h})T_{c}}{(T_{h}-T_{c})T_{w}}\equiv\epsilon_{c},
\end{equation}
where $\epsilon_{c}$ is upper bound on the COP for the heat-driven QAR operating between three independent thermal baths, which is nothing but the Carnot refrigerator efficiency.

	\begin{figure}[t]
		\centering
		\includegraphics[width=0.45\textwidth]{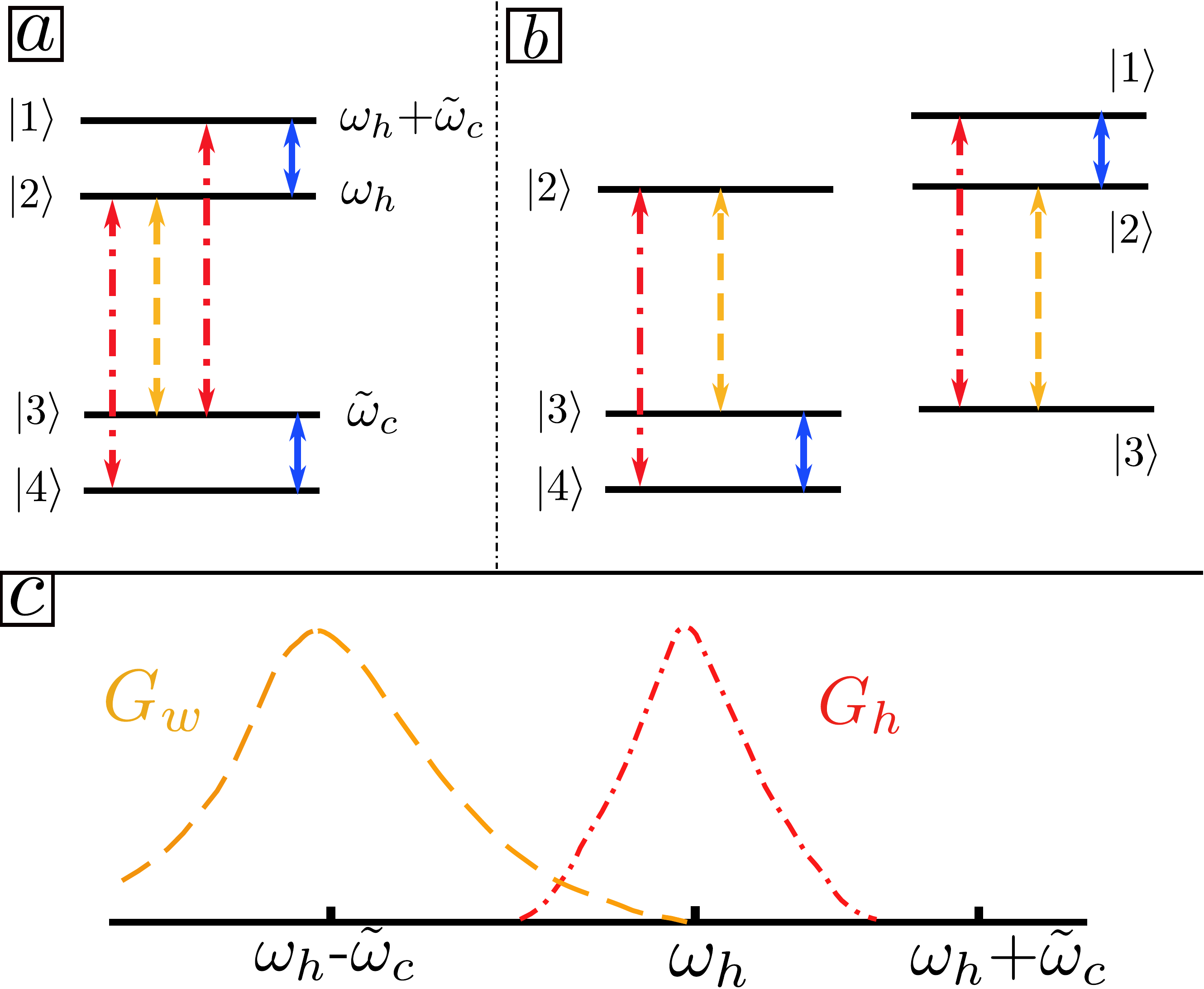}
		\caption{(Color online) Panel (a) shows the dressed energy levels of the coupled TLS and the arrows describe the possible transitions induced by the thermal baths whose spectral densities are presented in (c). The orange dashed, red dot-dashed and blue solid arrows illustrate the transitions induced by the work-like reservoir, hot and cold baths, respectively. (b) The qutrit breakup of our coupled qubits QAR is presented, it shows that our QAR can be considered as two three-level refrigerators functioning in parallel. The two qutrits share the work transition $2\leftrightarrow 3$, and note that the resonance condition $\omega_{h} = \tilde{\omega}_{c} + \omega_{w}$ holds.  (c) Shows the spectrally separated work (orange dashed) and hot (red dot-dashed) baths spectra given by Eq.~(\ref{eq:BathFilter}).}
		\label{fig:qRefLevels}
	\end{figure} 
	\begin{figure}[t]
		\centering
		\includegraphics[width=0.40\textwidth]{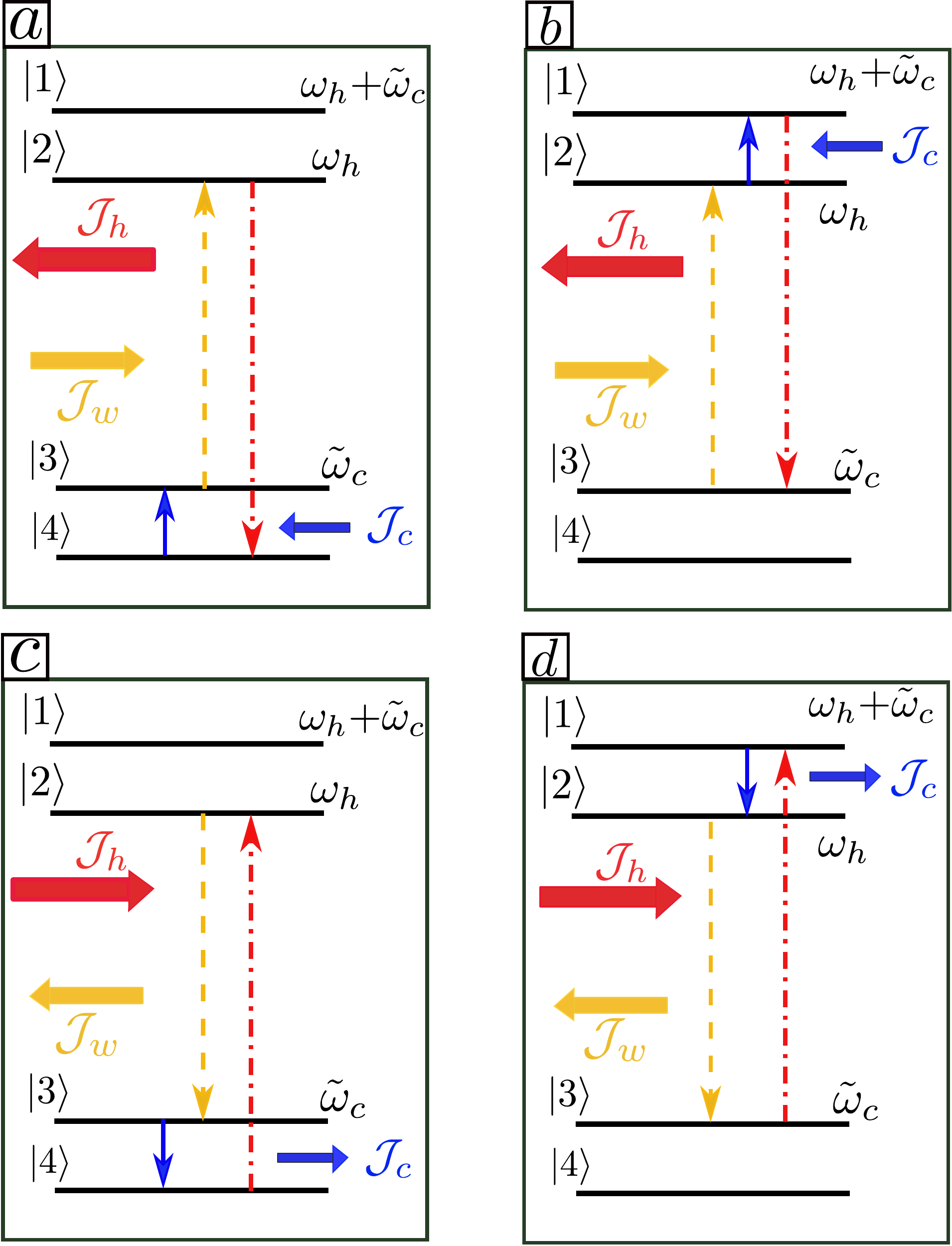}
		\caption{(Color online) The possible Raman cycles for the heat transfer.  Panels (a) and (b) show the only possible Raman cycles (4324) and (3213), respectively, the panels (c) and (d) show the inverses of these cycles. The orange dashed, red dot-dashed and blue solid arrows represent the transitions induced by the work-like reservoir, hot and cold baths, respectively. For $T_{w}>T_{h}>T_{c}$, within the cooling window, Raman cycles (4324) and (3213) are completed and their inverses are prohibited by the second law of thermodynamics, and only these two cycles are responsible for cooling.}
		\label{fig:qAbsorpMechanism}
	\end{figure} 

For the case of coupled TLS, transitions among the eigenstates of the Hamiltonian has shown in Fig.~\ref{fig:qRefLevels}(a), and details of which have given in Appendix~\ref{AppendixA}. Breakdown of TLS  model in two three-level (qutrit) subsystems is presented in Fig.~\ref{fig:qRefLevels}(b). Our model can be considered as two qutrit refrigerators functioning simultaneously that share a common work-like transition $|3\rangle\leftrightarrow|2\rangle$. Here, the resonance condition $\omega_{h}=\tilde{\omega}_{c}+\omega_{w}$ holds, and no transition is induced by more than one bath. Consequently, we have
\begin{equation}\label{eq:qeffratio}
\abs{\frac{\mathcal{J}_{\alpha}}{\mathcal{J}_{\beta}}}=\frac{\omega_{\alpha}}{\omega_{\beta}}.
\end{equation}
Similar result has also been shown for the case of ideal single-qutrit refrigerator in Ref.~\cite{PhysRev.156.343}, and we show that it also holds for our coupled qubit design presented in the Fig.~\ref{fig:qRefLevels}. In the high temperature limit $T_{w}\to\infty$, the steady-state heat currents~\cite{PhysRevE.85.061126} is given by
\begin{equation}\label{eq:new}
\mathcal{J}_{\alpha} = \omega_{\alpha} K, 
\end{equation}
(See Appendix~\ref{AppendixB} for details.) The Eq.~(\ref{eq:new}) also validates Eq.~(\ref{eq:qeffratio}) for the case of coupled TLS. However, in general, Eq.~(\ref{eq:qeffratio}) does not remain valid for the case of TLOS and OMS, because the resonance condition is not satisfied. This can be seen from the dressed Hamiltonians of TLOS and OMS, presented in Appendix.~\ref{AppendixA}. The presence of the $g^2$ dependent terms in Eqs.~(\ref{eq:acHami}) and ~(\ref{eq:omHami}) are responsible to move the system away from resonance condition.  However, in the weak-coupling regime $g\ll\omega_{c}$, we have confirmed numerically, Eq.~(\ref{eq:qeffratio}) is valid for both TLOS and OMS, as expected intuitively. 
 The COP is then evaluated and by using these equations and found to be
\begin{equation}\label{eq:qcarnoteff}
\epsilon= \frac{\mathcal{J}_{c}}{\mathcal{J}_{w}} = \frac{\tilde{\omega}_{c}}{\omega_{w}}.
\end{equation}
Using Eqs.~(\ref{eq:second}) and (\ref{eq:qeffratio}) we can write
\begin{equation}\label{eq:coolwindow}
\tilde{\omega}_{c}\leq\tilde{\omega}_{c, max}\equiv\frac{(T_{w}-T_{h})T_{c}}{(T_{h}-T_{c})T_{w}}\omega_{w},
\end{equation}
which gives the ``cooling window". Maximum COP is obtained when $\tilde{\omega}_{c}\to\tilde{\omega}_{c, max}$, for which the entropy production $\sigma\to 0$.

\subsection{Mechanism of cooling}
The spectral response functions of the work and hot baths are illustrated in the Fig.~\ref{fig:qRefLevels}(c), and we consider OSD for the cold bath.  With the selection of these spectra, transitions induced by all three baths are not overlapped. This makes our QAR model ideal so we can 
highlight the basic mechanism of cooling in our device. Additionally, for the case of spectrally separated thermal baths, Eq.~(\ref{eq:qcarnoteff}) holds and at resonance condition we can have efficiency arbitrarily close to the Carnot bound. Under the selection of different spectrally filtered bath spectra, we present our discussion in Sec.~\ref{sec:heatleak} and Appendix~\ref{app:FSQAR}.

	\begin{figure}[t]
		\centering
		\includegraphics[width=0.50\textwidth]{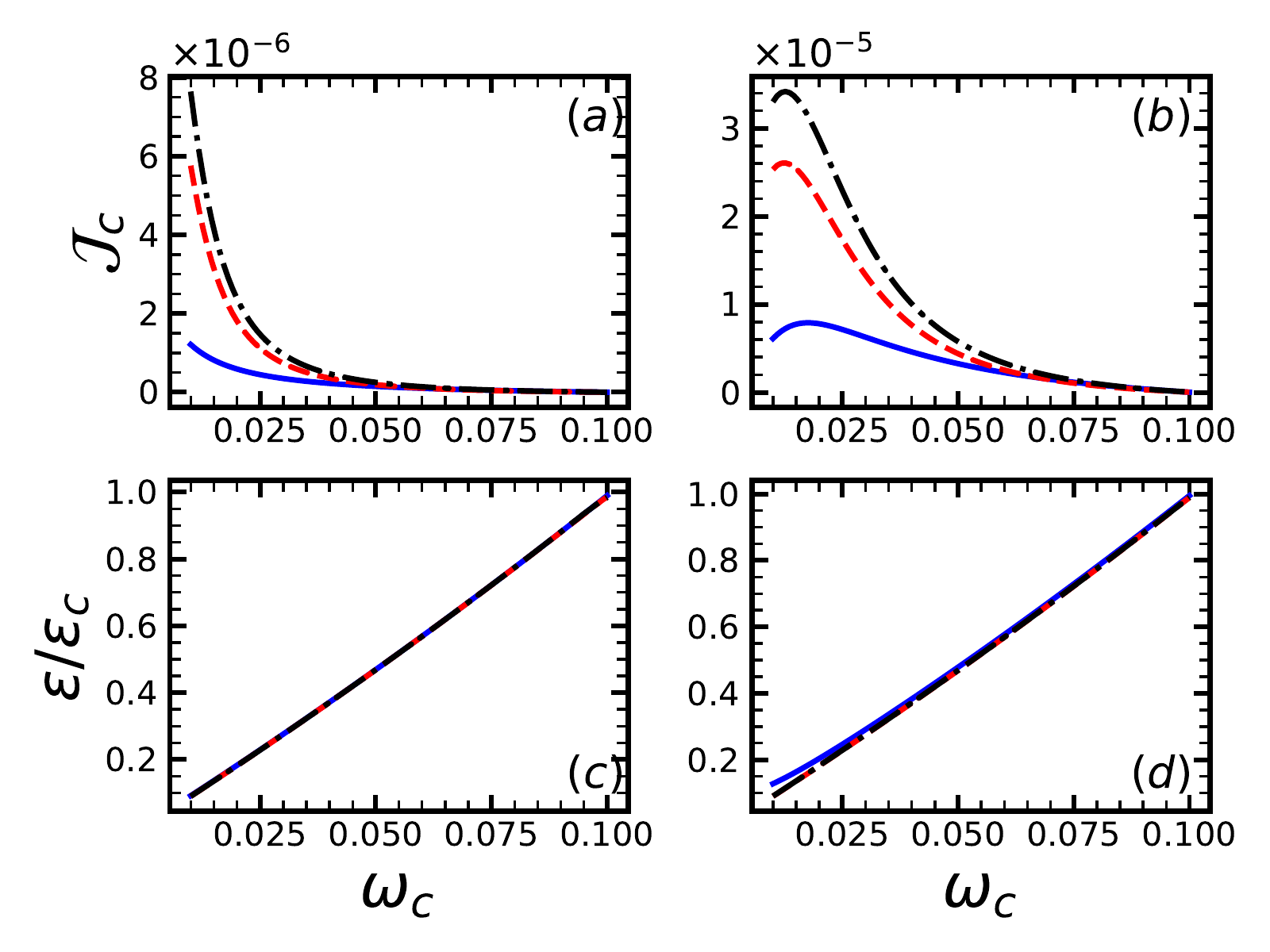}
		\caption{(Color online) (a), (b) The cooling power $\mathcal{J}_{c}$ and (c), (d) normalized coefficient of performance (COP) $\epsilon/\epsilon_{c}$ as a function of the subsytem B frequency $\omega_{c}$.  In all panels, solid, dashed, and dash-dotted lines are for TLS, TLOS and OMS, respectively. Parameters: (a), (c) $g=0.001$, (b), (d) $g=0.005$ and rest of the parameters are same for all panels. $\omega_{h}=1$, $\kappa_{w}=\kappa_{h}=\kappa_{c}=0.005$, $T_{w}=0.75$, $T_{h}=0.50$ and $T_{c}=0.125$. All the parameters are scaled with subsystem A frequency $\omega_{h}/2\pi=10$ GHz.}
		\label{fig:qeffec}
	\end{figure} 

For the selection of the bath spectra in Fig.~\ref{fig:qRefLevels}(c), only two global Raman cycles (4324)  [here (4324) means the sequence of transitions $|4\rangle \to |3\rangle \to |2\rangle \to |4\rangle$, etc.] and (3213) and their inverses are possible as illustrated in Fig.~\ref{fig:qAbsorpMechanism}. If the cycles in Figs.~\ref{fig:qAbsorpMechanism}(a) and \ref{fig:qAbsorpMechanism}(b) dominate, we have refrigeration effect, and we refer to these cycles as ``cooling cycles". However, if the inverse (`heating') cycles in Figs.~\ref{fig:qAbsorpMechanism}(c) and \ref{fig:qAbsorpMechanism}(d) are more favorable, then we get the heating effect. To decide which cycles are more likely, one needs to calculate the entropy production using the second law of thermodynamics. For example, using Eq.~(\ref{eq:second}), the entropy production for the cooling cycle in Fig.~\ref{fig:qAbsorpMechanism}(a) yields 
\begin{equation}
\sigma = - \frac{\tilde{\omega}_{c}}{T_{c}}- \frac{\omega_{w}}{T_{w}}+\frac{\omega_{h}}{T_{h}},
\end{equation}
so that the entropy production of this cycle $(4324)$ becomes $\sigma>0$ for system parameters within the cooling window presented in Eq.~(\ref{eq:coolwindow}). On the contrary, the entropy production for the heating cycles in Figs.~\ref{fig:qAbsorpMechanism}(c) and in~\ref{fig:qAbsorpMechanism}(d) are negative for the same set of parameters. Consequently, cooling cycles dominate and we get refrigeration effect for the considered system parameters. Similar cooling and heating cycles can be
identified for the TLOS and OMS by the breakdown of the system
into many qutrit subsystems working collectively. For example, we show two cooling cycles in
Fig.~\ref{fig:TLOS-OMS} in Appendix~\ref{AppendixA}.
For the choice of parameters within the cooling window, such cooling cycles dominate over their inverse heating cycles.

	\begin{figure}[t]
	\hspace*{-0.50cm}
		\centering
		\includegraphics[width=0.5\textwidth]{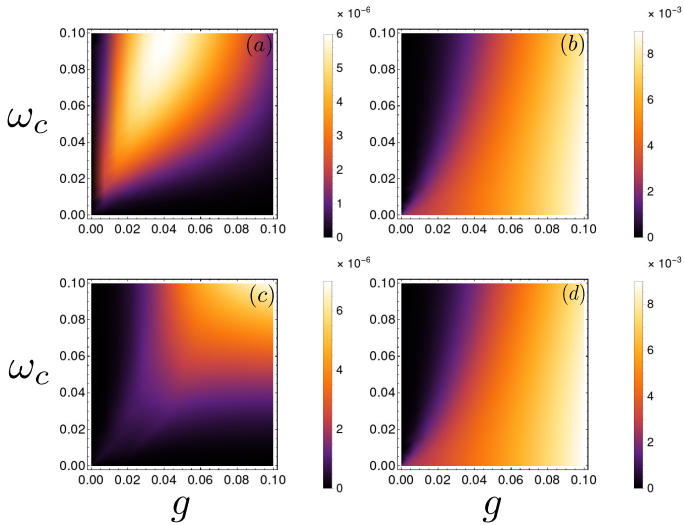}
		\caption{(Color online) (a) The cooling power $\mathcal{J}_{c}$, (b) total correlations $\mathcal{I}(\rho_{AB})$, (c) quantum correlations $\mathcal{D}(\rho_{AB})$, and (d) classical correlations $\mathcal{I}(\sigma_{AB})$ as a function of the $\omega_{c}$ and the coupling strength $g$, for the coupled TLS. Parameters: $\omega_{h}=1$, $\kappa_{w}=\kappa_{h}=\kappa_{c}=0.005$, $T_{w}=3$, $T_{h}=2$ and $T_{c}=1$.}
		\label{fig:R2}
	\end{figure} 

In Figs.~\ref{fig:qeffec}(a) and ~\ref{fig:qeffec}(c), the cooling power and scaled efficiency are plotted against normalized $\omega_{c}$, respectively, for the TLS, TLOS, and OMS in the weak coupling regime. Note that the efficiency curves are not far apart, because in the parameter regime $g\ll\omega_{c}$, for all three cases the efficiency becomes $\epsilon\approx\omega_{c}/\omega_{w}$. 
According to Eq.~(\ref{eq:qcarnoteff}), the COP can be varied arbitrarily close to the Carnot limit via controlling the frequency of subsystem B within the cooling window. In the limit $\epsilon\to\epsilon_{c}$, the entropy production $\sigma\to 0$ and the cooling power of QAR becomes zero, as expected for a reversible the Carnot refrigerator-type operation.  The cooling power and scaled efficiency are plotted against normalized $\omega_{c}$ for the TLS,TLOS, and OMS in Figs.~\ref{fig:qeffec}(b) and ~\ref{fig:qeffec}(d) for larger value of coupling strength $g$.  The cooling power $\mathcal{J}_{c}$ vanishes both at $\tilde{\omega}_{c}=0$ and $\tilde{\omega}_{c}=\tilde{\omega}_{c, max}$. For given system parameters, there exists a optimum value of $\tilde{\omega}_{c}$ for which the cooling power is maximum. This can be explained from the energy transitions shown in Fig.~\ref{fig:qAbsorpMechanism}. For optimal cooling effect following two conditions must hold, (i) the work bath must be hot enough to induce $|3\rangle\to|2\rangle$ transition; $T_{w}\gg\omega_{h}-\tilde{\omega}_{c}$, and (ii) the cold bath must have $T_{c}>\tilde{\omega}_{c}$. For given system parameters, at small values of $\tilde{\omega_{c}}$, condition (i) may not be satisfied, and for higher values of $\tilde{\omega_{c}}$, (ii) may not be true. Hence, the optimal cooling can be achieved when both of these conditions are satisfied.

\subsection{Hilbert space size, quantum correlations and cooling power}
We compare the cooling power $\mathcal{J}_{c}$ of three considered working mediums  in Figs.~\ref{fig:qeffec}(a) and~\ref{fig:qeffec}(b). The cooling power of the working mediums are found to be in the following order: $\mathcal{J}^{\text{OMS}}_{c}>\mathcal{J}^{\text{TLOS}}_{c}\gg\mathcal{J}^{\text{TLS}}_{c}$. This is because the size of Hilbert space for the coupled TLS is smallest among all three working mediums, and the Hilbert space dimensions should be considered as a thermodynamic resource~\cite{PhysRevE.89.042128, PhysRevE.94.032120}.

	\begin{figure}[t]
	\hspace*{-0.50cm}
		\centering
		\includegraphics[width=0.50\textwidth]{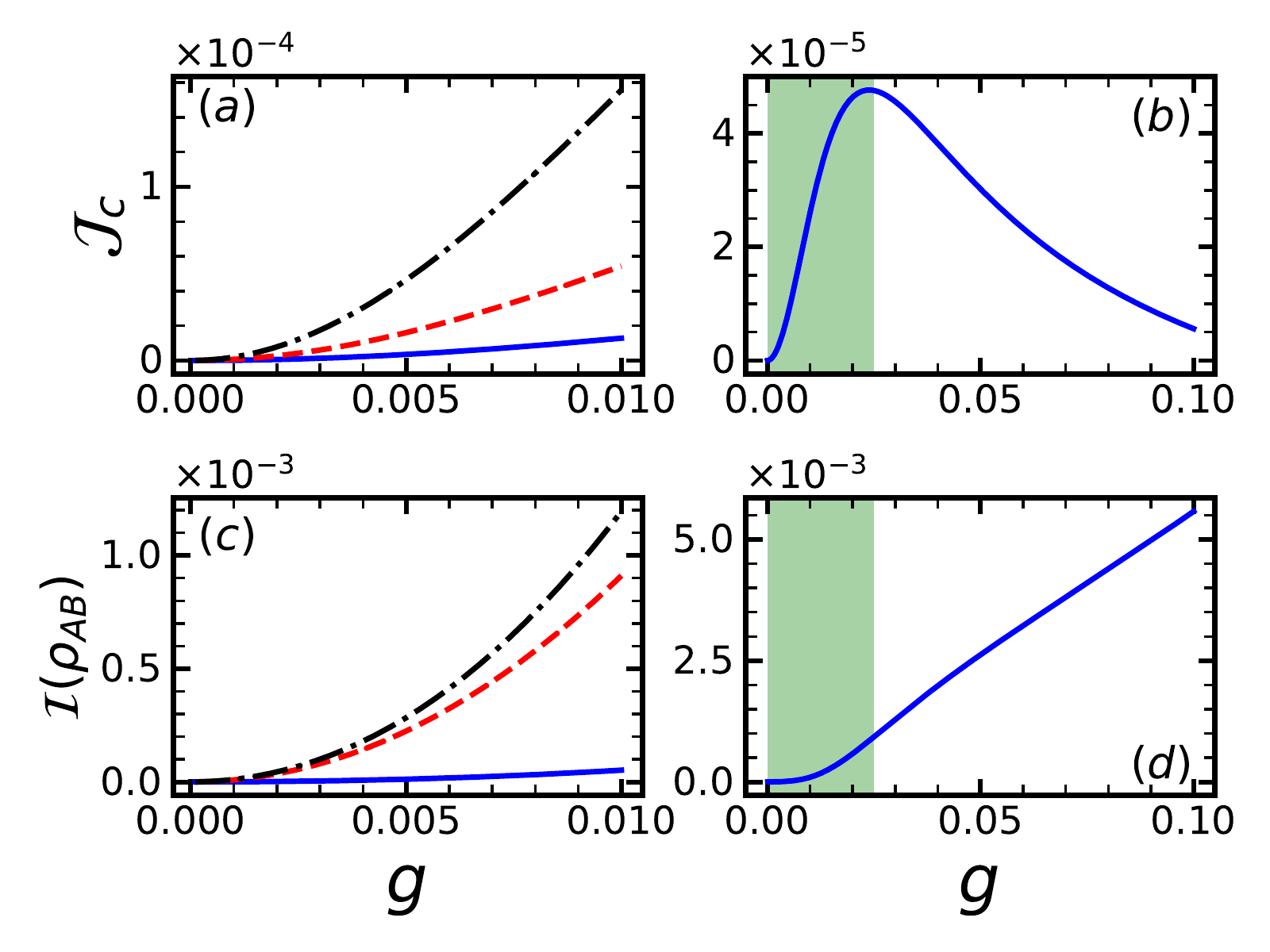}
		\caption{(Color online) (a), (b) The cooling power $\mathcal{J}_{c}$ and (c), (d) mutual information $\mathcal{I}({\rho_{AB}})$ between subsystems A and B as a function of coupling strength $g$.  In all panels, solid, dashed, and dash-dotted lines are for TLS, TLOS and OMS, respectively. Results in panels (a) and (c) are for weak coupling regime $\omega_{c}=0.1$, and in panels (b), (d) we have $\omega_{c}=0.025$ for the TLS, and shaded regions show the weak coupling regime. Parameters: $\omega_{h}=1$, $\kappa_{w}=\kappa_{h}=\kappa_{c}=0.005$, $T_{w}=3$, $T_{h}=2$ and $T_{c}=1$.}
		\label{fig:MutualInform}
	\end{figure} 

Now we explore the role of quantum correlations on the performance of our QAR. We find no entanglement between the subsystems A and B in the considered parameters regime. To explain this, we express the density matrix of the coupled TLS in the individual eigenstates of the qubits (Appendix~\ref{AppendixA}), given by
\begin{equation}
  \rho_{AB}=
  \begin{pmatrix}
    \rho_{11} & \rho_{12} & 0 & 0 \\
    \rho_{12} & \rho_{22} & 0 & 0 \\
    0 & 0 & \rho_{33} & \rho_{34} \\
    0 & 0 & \rho_{34} & \rho_{44}
  \end{pmatrix}.
  \label{eq:matrix}
\end{equation}
For the considered parameters, the least eigenvalue of the partial-transpose of the density matrix $\rho_{AB}$ remains positive; $\rho_{11}+\rho_{22}-((\rho_{11}-\rho_{22})^2+4\rho_{12}^{2})^{1/2}>0$. Consequently, $\rho_{AB}$ is not entangled following the positivity-of-the-partial-transpose separability criterion~\cite{PhysRevLett.77.1413}. To quantify the quantum correlations between the coupled TLS, that are beyond entanglement-separability paradigm, we consider the quantum discord, which is defined as~\cite{PhysRevLett.88.017901, Henderson_2001}
\begin{equation}\label{eq:discord}
D(\rho_{AB}) := \mathcal{I}(\rho_{AB}) - \mathcal{I}(\sigma_{AB}).
\end{equation} 
Where, $\mathcal{I}(\rho_{AB})$ is the quantum mutual information which quantifies the total correlations in the system and given by
\begin{equation}\label{eq:mutual}
\mathcal{I}(\rho_{AB}) = S_{\rho_{A}} + S_{\rho_{B}} - S_{\rho_{AB}}.
\end{equation}
$S_{\rho_i}= -\text{Tr}[\rho_i\text{log}\rho_i]$ is the von Neumann entropy of the reduced density matrix of the subsystem $i$. In Eq.~(\ref{eq:discord}), $\mathcal{I}(\sigma_{AB})$ quantifies the classical correlations in the system, where $\sigma_{AB}$ is the
resultant state of minimally disturbing projective measurement on qubit B of $\rho_{AB}$~\cite{PhysRevLett.88.017901, Henderson_2001}.  To look for the possible role of correlations on the maximization of the cooling power, we plot the cooling power $\mathcal{J}_{c}$, 
quantum correlations $D(\rho_{AB})$, classical correlations $\mathcal{I}(\sigma_{AB})$, and total correlations $\mathcal{I}(\rho_{AB})$ in Fig.~\ref{fig:R2}. The quantum correlations appear to be much weaker than the classical correlations. The region of the maximum cooling power does not coincide with the quantum or classical correlations. Similar results are obtained for COP at maximum cooling power $\epsilon_{*}$. We find zero discord, $D(\rho_{BA}) = 0$, when
minimally disturbing projective measurement is performed on
qubit A.
	\begin{figure}[t]
	\hspace*{-0.50cm}
		\centering
		\includegraphics[width=0.50\textwidth]{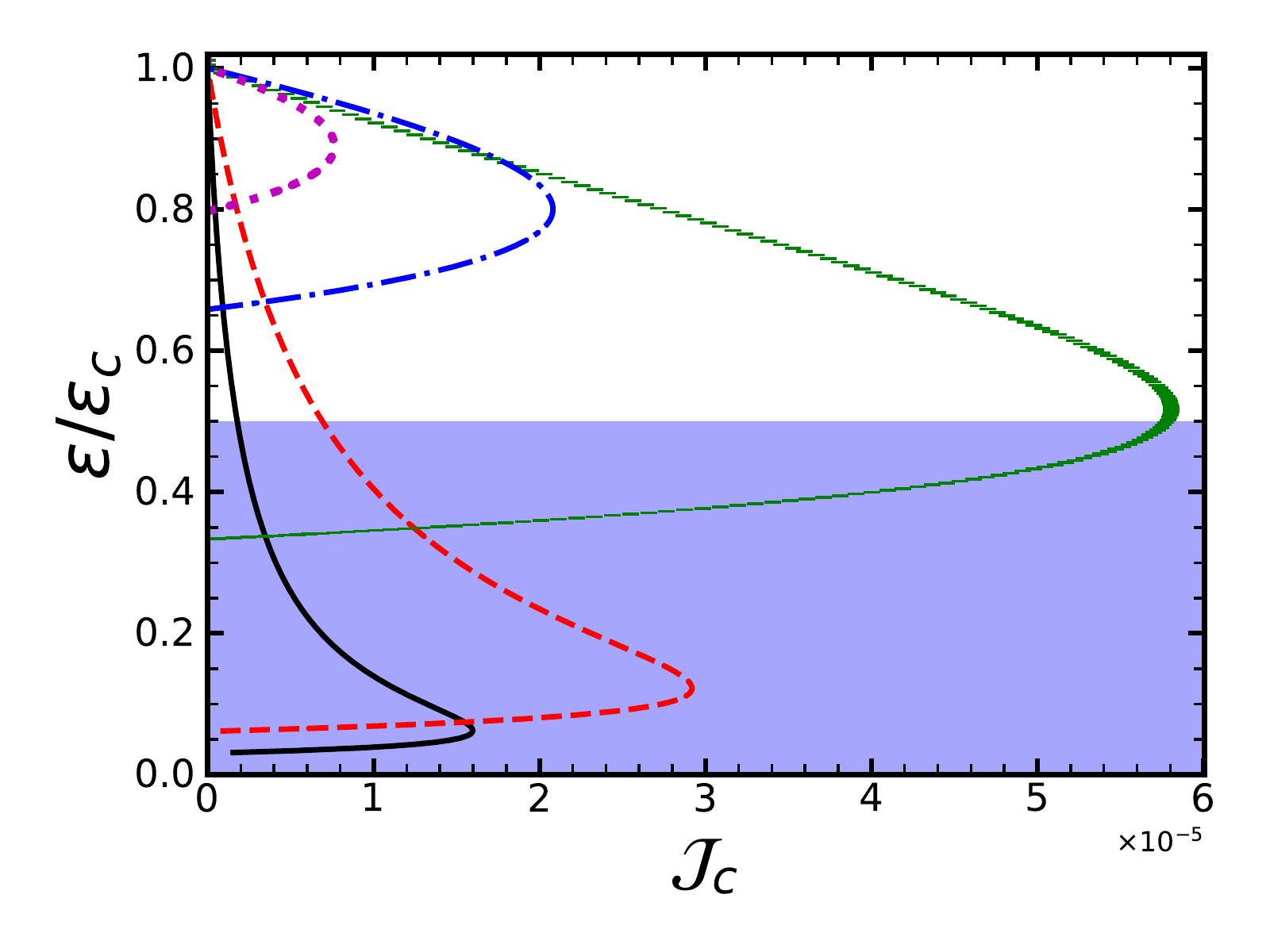}
		\caption{(Color online) Parametric plot of cooling power $\mathcal{J}_{c}$ vs. normalized coefficient of performance (COP) $\epsilon/\epsilon_{c}$ for different values of the coupling strength $g$. Solid, dashed, horizontal dashed, dash-dotted and dotted lines are for $g={0.005, 0.01, 0.05, 0.09, and ~0.11}$, respectively. The rest of the parameters are the same as in Fig.~\ref{fig:qeffec}.
		The shaded area shows the efficiency at maximum power $\epsilon_{*}$ bound for one dimensional bosonic reservoirs given in Eq.~(\ref{eq:ebound}).}
		\label{fig:COPMP}
	\end{figure} 

The computation of the quantum discord is numerically intractable, as the computing time grows exponentially with the size of the Hilbert space~\cite{Huang_2014}. Except for Gaussian states~\cite{PhysRevLett.105.020503}, there is no known
method of quantum discord calculation for continuous
variable states.
Due to the nonlinear nature of
optomechanical coupling, our OMS may not have a 
Gaussian steady-state in general.
Accordingly, while we report both the quantum discord and
quantum mutual information for coupled TLS in Fig.~\ref{fig:R2},
we are limited to present only the quantum mutual information
for OMS in Fig.~\ref{fig:MutualInform}.  
In Figs.~\ref{fig:MutualInform}(a) and~\ref{fig:MutualInform}(c), we plot the cooling power $\mathcal{J}_{c}$ and quantum mutual information $\mathcal{I}(\rho_{AB})$ as a function of coupling strength $g$ in the weak coupling limit. For the coupled TLS, we plot the cooling power $\mathcal{J}_{c}$ and mutual information $\mathcal{I}(\rho_{AB})$ in Figs.~\ref{fig:MutualInform}(b) and \ref{fig:MutualInform}(d), respectively. Here both the weak and strong coupling regimes are considered. 
Our numerical observations based upon 
the quantum discord for coupled TLS (Fig.~\ref{fig:R2}) and the quantum mutual information for all three working mediums (Fig.~\ref{fig:MutualInform}), implies that the maximum cooling power $\mathcal{J}^{*}_{c}$ is not associated with the maximum of total correlations in the system (c.f. with the Ref.~\cite{PhysRevE.87.042131} which also supports our results).

\subsection{Efficiency at maximum power}
	\begin{figure}[t]
	\hspace*{-0.50cm}
		\centering
		\includegraphics[width=0.50\textwidth]{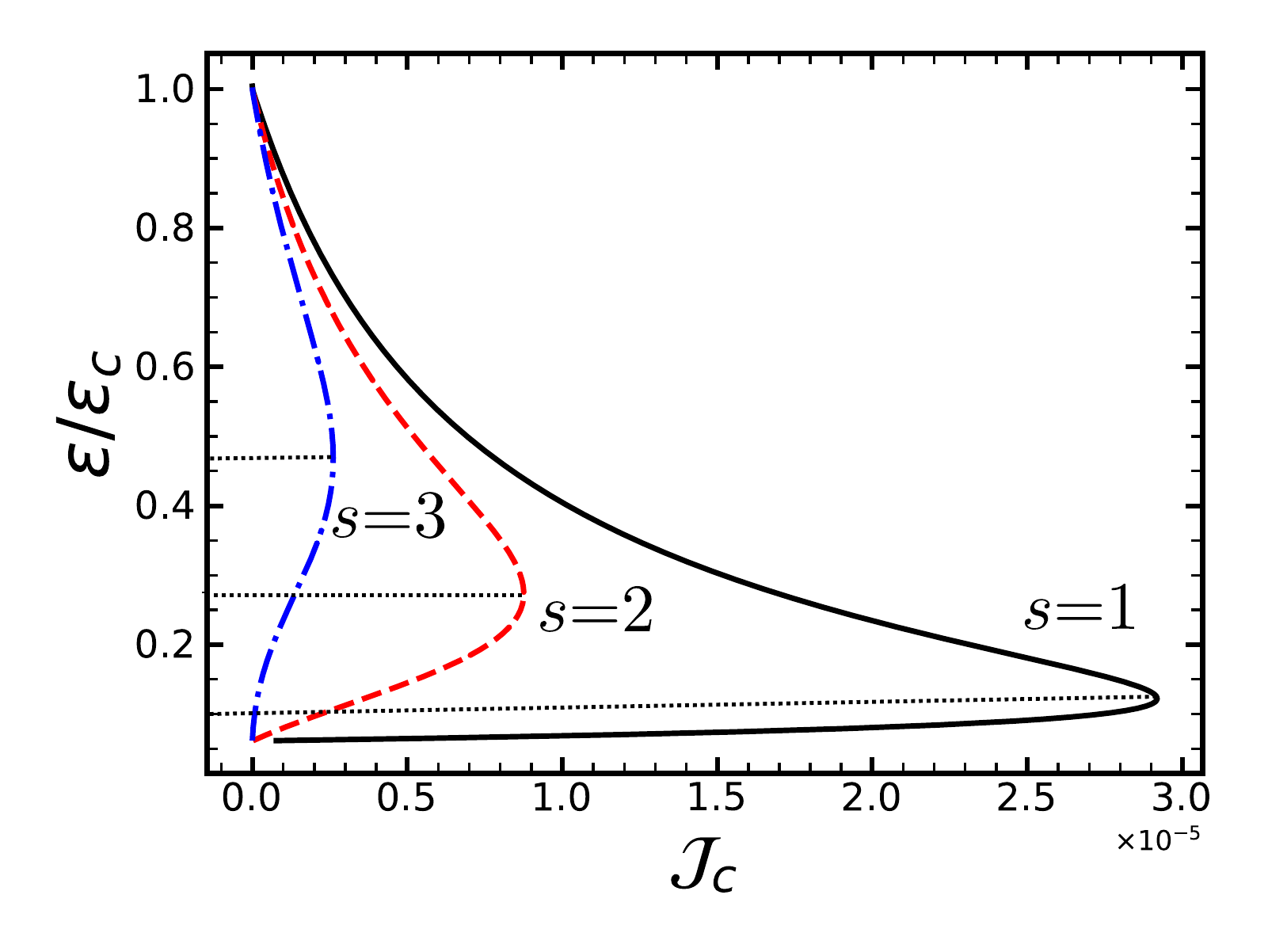}
		\caption{(Color online) Parametric plot of cooling power $\mathcal{J}_{c}$ vs. normalized coefficient of performance (COP) $\epsilon/\epsilon_{c}$ for different bath spectral densities. Solid, dashed, and dash-dotted lines are for $s={1, 2, 3}$, respectively, for details see Eq.~(\ref{eq:BSD}). The dotted lines show COP at maximum cooling power.
		Parameters: $T_{w}=3$,  $T_{h}=2$, $T_{c}=1$, $\omega_{h}=1$, $\kappa_{w}=\kappa_{h}=\kappa_{c}=0.005$, and $g=0.01$.
		}
		\label{fig:Refree1}
	\end{figure} 
In addition to the Carnot bound, a more practical bound for quantum thermal machines is efficiency at maximum power $\epsilon_{*}$. At the limit of the Carnot bound, the cooling power vanishes due to the reversibility of the ideal QAR as shown in Figs.~\ref{fig:qeffec}(c) and ~\ref{fig:qeffec}(d). In Ref.~\cite{Correa2014}, it has been argued that the efficiency at maximum cooling power of any ideal QAR composed of elementary qutrit fridges is tightly upper bounded by
\begin{equation}\label{eq:ebound}
\epsilon_{*} \leq \frac{d_{c}}{d_{c} +1}\epsilon_{c},
\end{equation}
where $d_{c}$ is the spatial dimensionality of the cold bath. We show that one can surpass this bound in a strong coupling regime, contrary to previously proposed models~\cite{PhysRevLett.105.130401, PhysRevE.85.061126, PhysRevE.87.042131}. However, in the case of TLOS and OMS, Eq.~(\ref{eq:master}) is only valid in weak coupling regime
$g^{2}\langle\tilde{b}^{\dagger}\tilde{b}\rangle\ll\omega^{2}_{c}$. Therefore, in the rest of the paper, we only consider coupled TLS in the strong coupling regime with one dimensional OSD of the baths.
	\begin{figure}[t]
	\hspace*{-0.50cm}
		\centering
		\includegraphics[width=0.50\textwidth]{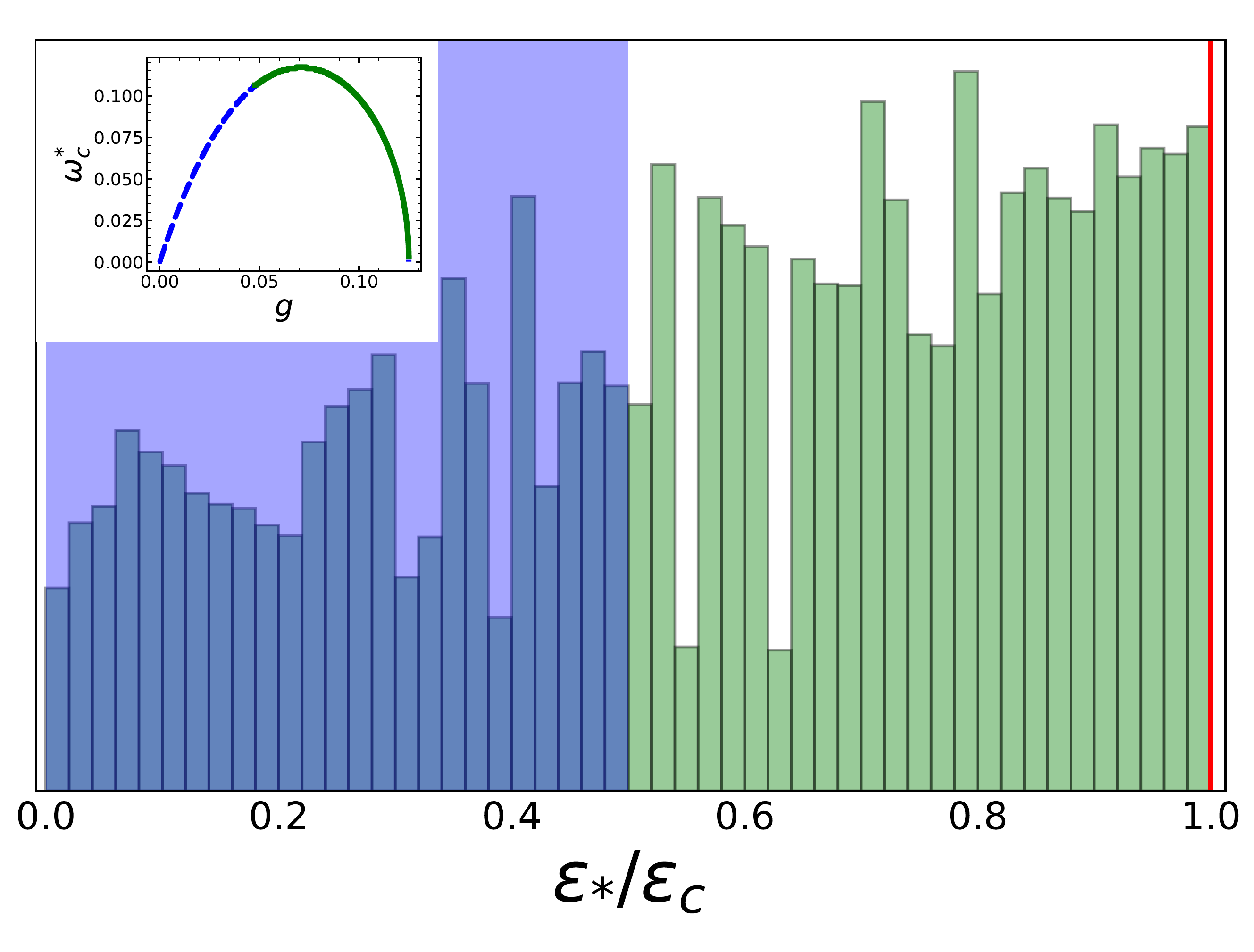}
		\caption{(Color online) Histogram of efficiency at maximum cooling power $\epsilon_{*}/\epsilon$ for the coupled two-level system. The shaded area shows the efficiency at maximum power $\epsilon_{*}/\epsilon$ bound for one-dimensional bosonic reservoirs given in Eq.~(\ref{eq:ebound}). The solid line on the right of the plot shows the Carnot bound. We have considered  $5\times 10^{6}$ randomly selected system parameters $\{T_{\alpha}, \kappa_{\alpha}, g\}$ sets within the cooling window. We fix the value of subsystem A frequency; $\omega_{h}=1$. For every choice of one of these system parameters set, we vary $\omega_{c}$ to find the optimum value that maximized the cooling power $\mathcal{J}_{c}$. It can be seen that we can beat the bound of Eq.~(\ref{eq:ebound}) and even can reach arbitrarily close to the Carnot limit.  Inset shows the values of coupling strength $g$ and subsystem B frequency $\omega^{*}_{c}$ obtained at $\epsilon_{*}$. Solid and dashed lines are associated with above and below the bound given in Eq.~(\ref{eq:ebound}). It is clear that the efficiency at maximum power can be reached close to the Carnot limit operating at $g\gg\omega^{*}_{c}$ and $\omega^{*}_{c}\to 0$, which makes cooling power $\mathcal{J}^{*}_{c}\to 0$.}
		\label{fig:COPMPb}
	\end{figure} 
In Fig.~\ref{fig:COPMP}, we plot the normalized COP versus cooling power for different values of the coupling strength $g$. The shaded region shows the efficiency at maximum power bound $\epsilon_{*}$ presented in Eq.~(\ref{eq:ebound}). The increment in the coupling strength $g$ results in the increment of $\epsilon_{*}$, and one can surpass the bound given in Eq.~(\ref{eq:ebound}) for higher values of $g$. The efficiency at maximum cooling power can be made arbitrarily close to the Carnot bound $\epsilon_{*}\to\epsilon_{c}$, though the power gradually approaches to zero as the efficiency approaches the Carnot bound.
The bound $\epsilon_{*}$ depends on the dimensionality of the cold bath, with the increase in the dimensions of the cold bath, this bound goes higher. For $d$-dimensional bath the power spectrum is $G(\tilde{\omega_{c}})\propto\tilde{\omega}^{d}_{c}$, and in our case,  the presence of $c=\omega_{c}/\tilde{\omega}_{c}$  coefficient in Eq.~(\ref{eq:master}), effectively changes the dimensionality of the cold bath. Consequently, 
the increase in $\epsilon_{*}$ as compared to previously proposed models is due to the non-linear coupling between the subsystem A and B. In Fig.~\ref{fig:Refree1}, parametric plot of cooling power $\mathcal{J}_{c}$ as a function of normalized coefficient of performance $\epsilon/\epsilon_{c}$ is shown for bath spectrums different than Ohmic ($s\ne 1$), for details see Eq.~(\ref{eq:BSD}). The qualitative behavior of cooling power and coefficient of performance remains the same, however, efficiency at maximum power $\epsilon_{*}$  becomes quantitatively different.

Here we like to point out  that, for $g=N\omega_{c}$, where $N$ is a real number, the coefficient $c$ (Eq.~(\ref{eq:master}))  no longer depends on $\omega_{c}$. Then it is not possible to beat the bound in Eq.~(\ref{eq:ebound}) even for arbitrarily large coupling strengths $g$, as elaborated in Appendix~\ref{app:wcEff}. In Fig.~\ref{fig:COPMPb},  we fix the system parameters such that $T_{w}>\omega_{w}$ and $T_{h}\gg T_{c}$. For every value of $g$, we find the optimum value of $\omega_{c}$ at which we obtain $\epsilon_{*}$. The $\omega^{*}_{c}$ scales quadratically with the coupling strength $g$, and to operate at $\epsilon_{*}\to\epsilon_{c}$ we should have $g\gg\omega_{c}$ and $\omega_{c}\to 0$.
 The histogram of scaled efficiency at maximum cooling power $\epsilon_{*}/\epsilon$ of coupled TLS for randomly selected system parameters $\{T_{\alpha}, \kappa_{\alpha}, g\}$ has been presented in Fig.~\ref{fig:COPMPb}. All system parameters are considered within the cooling window such that the Eq.~(\ref{eq:master}) remains valid. It is clear that $\epsilon_{*}$ can be made arbitrarily close to   the Carnot limit. To achieve the efficiency at maximum power close to the Carnot bound, the sufficient condition on system parameters is $\omega_{w}\ll T_{w}$, $T_{h}\gg T_{c}$, and $g\gg\omega_{c}$.

\section{Heat leaks: non-ideal QAR}\label{sec:heatleak}
	\begin{figure}[t]
	\hspace*{-0.50cm}
		\centering
		\includegraphics[width=0.50\textwidth]{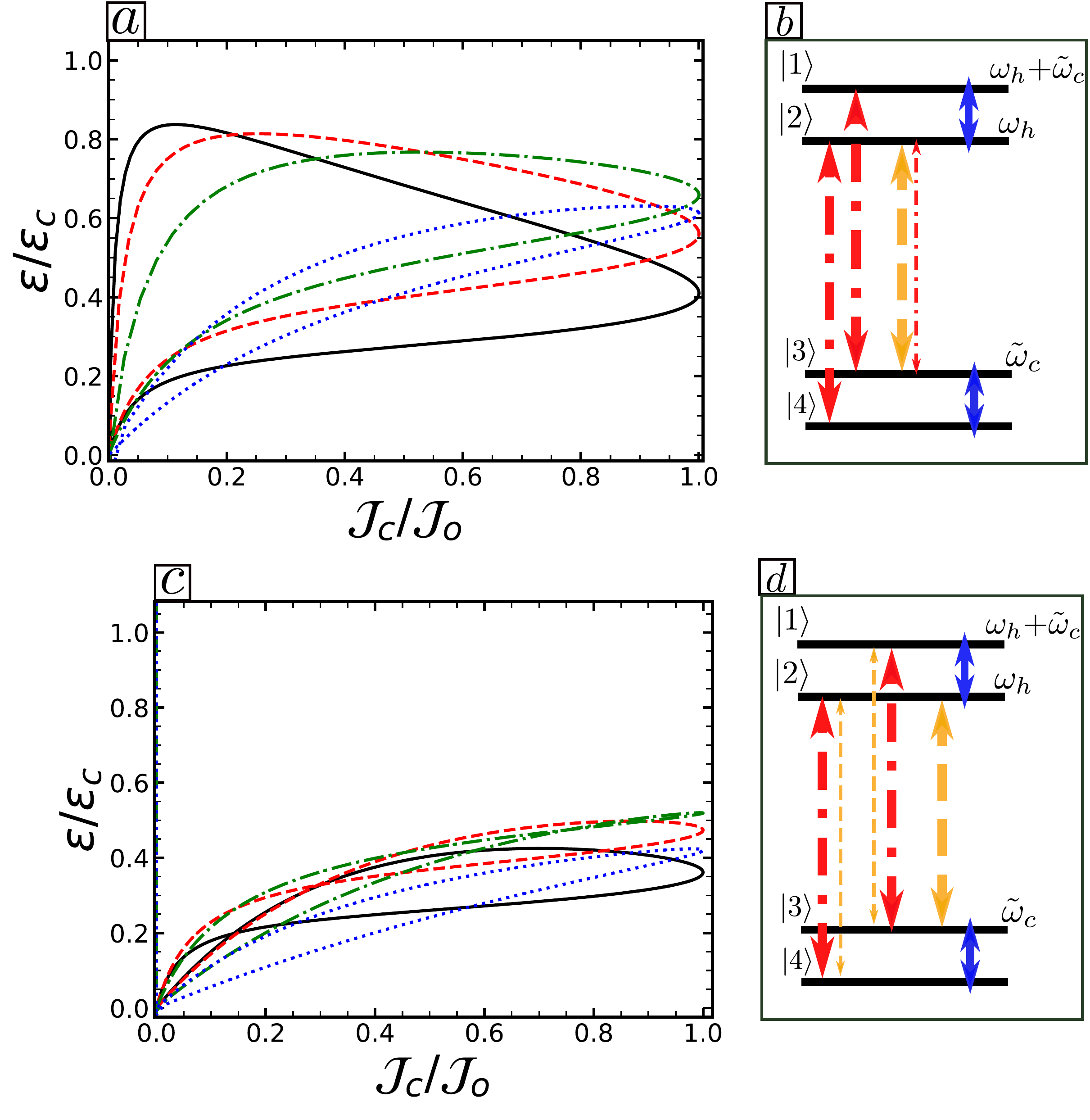}
		\caption{(Color online) (a), (c) Parametric plot of normalized cooling power $\mathcal{J}_{c}/\mathcal{J}_{0}$ vs. normalized coefficient of performance (COP) $\epsilon/\epsilon_{c}$ for different values of the coupling strength $g$. In panel (b), $\omega_{w}=\omega_{h}-\tilde{\omega_{c}}$ transition is induced by both work and hot baths. The weight of the arrows shows the transition rates. In panel (d), $\omega_{h}$ transitions are induced by both hot and work baths. In (c) and (d), solid, dashed, dash-dotted arrows represent the transitions induced by cold, work and hot baths, respectively. In panels (a) and (c), solid, dashed, dash-dotted and dotted lines are for $g={0.04, 0.06, 0.08, and ~0.10}$, respectively. Rest of the parameters are same as in Fig.~\ref{fig:qeffec}. }
		\label{fig:heatleak}
	\end{figure} 

Thermal devices are normally characterized by their efficiency vs. power curves. The ideal QAR has open characteristic curves as shown in Fig.~\ref{fig:COPMP}. It indicates that in the reversible limit, the COP saturates to the Carnot bound at the cost of the vanishing of cooling power. However, more realistic models of QAR suffer through heat leaks and internal dissipation that deteriorates the COP of the fridge~\cite{PhysRevE.92.032136}. For the case of these non-ideal QARs, the characteristic curves for power vs. efficiency are closed. 

Here we consider a more realistic scenario compared to previous sections. We model the irreversible process in our device by allowing one or more energy transitions to be induced by two or more thermal baths.  This can happen if the thermal baths coupled to subsystem A are not spectrally well separated, instead they overlap on one or more transitions.
For example, if the work transition ($\omega_{w}$) is induced by both work and hot baths, it can be a source of irreversibility~\cite{PhysRevE.64.056130}. In our model, this is shown in Fig.~\ref{fig:heatleak}(b) and corresponding efficiency vs. power plot has been shown in Fig.~\ref{fig:heatleak}(a) for different values of coupling strength $g$.  As expected, we have closed efficiency vs. power curves, and with the increase in the coupling strength the irreversibility increases. The coupling of a hot bath with work transition is parasitic, which introduces the heat leaks in the device, hence the irreversibility. Due to this parasitic coupling, in addition to cycles shown in Fig.~\ref{fig:qAbsorpMechanism}, three additional cycles and their inverses are possible; $4\xrightarrow{T_{c}}3\xrightarrow{T_{h}}2\xrightarrow{T_{h}}4$, $3\xrightarrow{T_{h}}2\xrightarrow{T_{c}}1\xrightarrow{T_{h}}3$,  $2\xrightarrow{T_{w}}3\xrightarrow{T_{h}}2$, and this leads to thermal short-circuit. In Fig.~\ref{fig:heatleak}(d), $\omega_{h}$ transition is coupled with both work and a hot bath, and corresponding irreversible efficiency vs. cooling power curves are shown in Fig.~\ref{fig:heatleak}(c). 
Qualitatively similar results are obtained for TLOS and OMS in the weak coupling regime.

\section{Conclusions}\label{sec:conclusions}
We have proposed a model of quantum absorption refrigerator based upon two-body interaction instead of widely used three-body interaction. The model is analyzed for a working medium that consists of two coupled qubits or two harmonic oscillators or one qubit and one harmonic oscillator. A global master equation is employed to study the dynamics of QAR in both the weak and strong coupling regime of the working medium. The refrigeration is explained by Raman transitions between the dressed energy levels of the interacting working medium. By reservoir engineering and selection of suitable system parameters, the net effect of these cycles results in the transfer of heat from cold to a hot bath. For all three working mediums, the coefficient of performance arbitrarily close to the Carnot bound has been achieved at which the cooling power vanishes. In the strong coupling regime, the efficiency at maximum power has saturated the Carnot bound (the power gradually goes to zero at the Carnot point), which is not possible with previously proposed models of QARs. We investigated a more realistic model of our QAR that suffers from heat leaks.
 We use the experimentally realizable system parameters in our numerical simulations.
 
Our proposal of QAR presents a versatile experimental platform for its realization.
For example, it can be implemented in circuit QED architectures; for coupled qubits~\cite{PhysRevE.99.042121}, for atom-cavity model~\cite{PhysRevE.90.022102} and for the optomechanical system as working medium~\cite{PhysRevLett.120.227702}. Unlike previously proposed optomechanical QAR models
~\cite{Mitchison_2016,PhysRevLett.108.120602} which require three-body interaction, our model require only two-body interaction. From the conceptual point of view, we provide a simple interpretation of the mechanism of cooling in non-linear optomechanical systems. Compared to the previous two-body QARs proposals~\cite{Correa2014, PhysRevE.92.012136, PhysRevE.85.061126,doi:10.1080/00107514.2019.1631555}, in our model, using reservoir engineering, both the efficiency and efficiency at maximum power can be made very close to the Carnot limit.
\\

{\bf Acknowledgements:} M.T.N acknowledges O. Pusuluk for the fruitful discussions, and C. L. Latune for the useful comments on the the manuscript.

\appendix

\section{\label{AppendixA} Eigenoperators and dressed Hamiltonians}
\label{app:DressStates}
	\begin{figure}[t]
		\centering
		\includegraphics[width=0.40\textwidth]{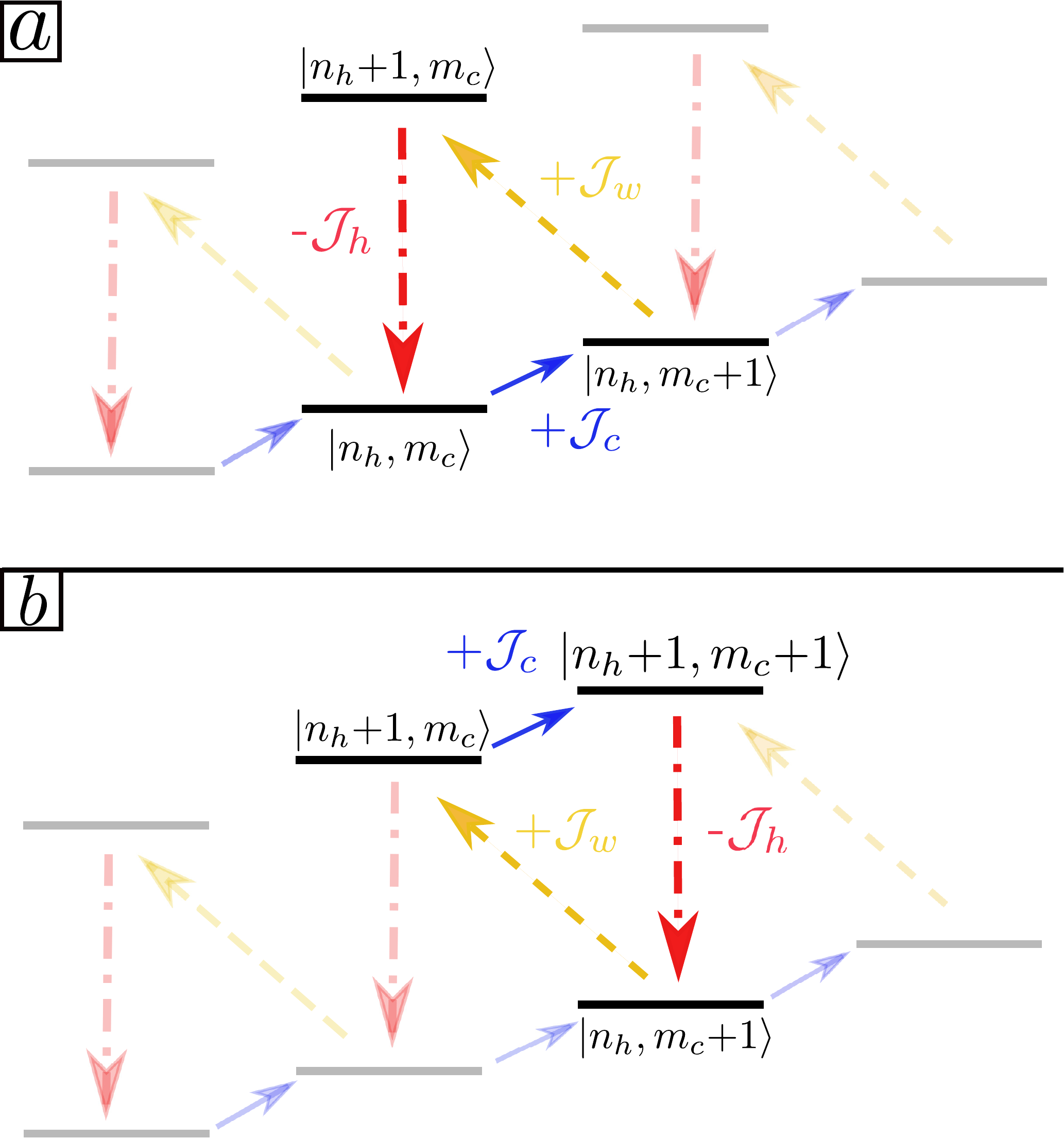}
		\caption{(Color online) Examples of possible cooling cycles in two-level-oscillator and optomechanical systems.  $|n_{h}, m_{c}\rangle$ are the dressed states, where $n_{h}=0,1$ represent the ground and excited state of the two-level system for TLOS, and $n_{h}=0,1,2,...$ give the photon number for OMS. Furthermore, $m_{c}=0,1,2,...$ for both TLOS and OMS. The orange dashed, red dot-dashed and blue solid arrows represent the transitions induced by the work-like reservoir, hot and cold baths, respectively. For $T_{w}>T_{h}>T_{c}$, within the cooling window, these cycles (cooling) are completed and their inverses (heating) are prohibited by the second law of thermodynamics.}
		\label{fig:TLOS-OMS}
	\end{figure} 

Here we present the explicit form of eigenoperators and diagonalized Hamiltonians for all three cases.

(i) {\it{Coupled two two-level system}}: by applying the unitary transformation given in 
Eq.~(\ref{eq:unitary}) on the Hamiltonian of Eq.~(\ref{eq:Hamsys}) we get
\begin{equation}
\tilde{H}_{qq} = \frac{\omega_{h}}{2}\tilde{\sigma}_{h}^{z}+\frac{\tilde{\omega_{c}}}{2}\tilde{\sigma}_{c}^{z},
\end{equation}
where the transformed Pauli operators read
\begin{eqnarray}
{\tilde{\sigma}}_{h}^{z}&=& U \hat{\sigma}_{h}^{z}U^\dagger= \hat{\sigma}_{h}^{z},\label{eq:transforme}\\
{\tilde{\sigma}}_{c}^{z}&=& U \hat{\sigma}_{c}^{z}U^\dagger= \cos\theta\hat{\sigma}_{c}^{z} - \sin\theta\hat{\sigma}_{h}^{z}\hat{\sigma}_{c}^{x}.\label{eq:transforme}
\end{eqnarray}

The dressed Hamiltonian has eigenstates $|i\rangle$, $i=1,2,3,4$ and described in terms of individuals eigenstates of the qubits as
\begin{eqnarray}\label{eq:eigenstate1}
\ket{1} = \cos\frac{\theta}{2}\ket{++} - \sin\frac{\theta}{2}\ket{+-} \nonumber \\ 
\ket{2} = \sin\frac{\theta}{2}\ket{++} + \cos\frac{\theta}{2}\ket{+-} \nonumber \\
\ket{3} = \cos\frac{\theta}{2}\ket{-+} + \sin\frac{\theta}{2}\ket{--} \nonumber \\
\ket{4} = \cos\frac{\theta}{2}\ket{--} - \sin\frac{\theta}{2}\ket{-+}\label{eq:eigenstate4},
\end{eqnarray}  
the corresponding eigenvalues are $\omega_{1} = (\omega_{h} + \tilde{\omega}_{c})/2$,  $\omega_{2} = (\omega_{h} - \tilde{\omega}_{c})/2$, $\omega_{3} = (-\omega_{h} + \tilde{\omega}_{c})/2$, and $\omega_{4} = (-\omega_{h} - \tilde{\omega}_{c})/2$, respectively. 

	\begin{figure}[t]
		\centering
		\includegraphics[width=0.45\textwidth]{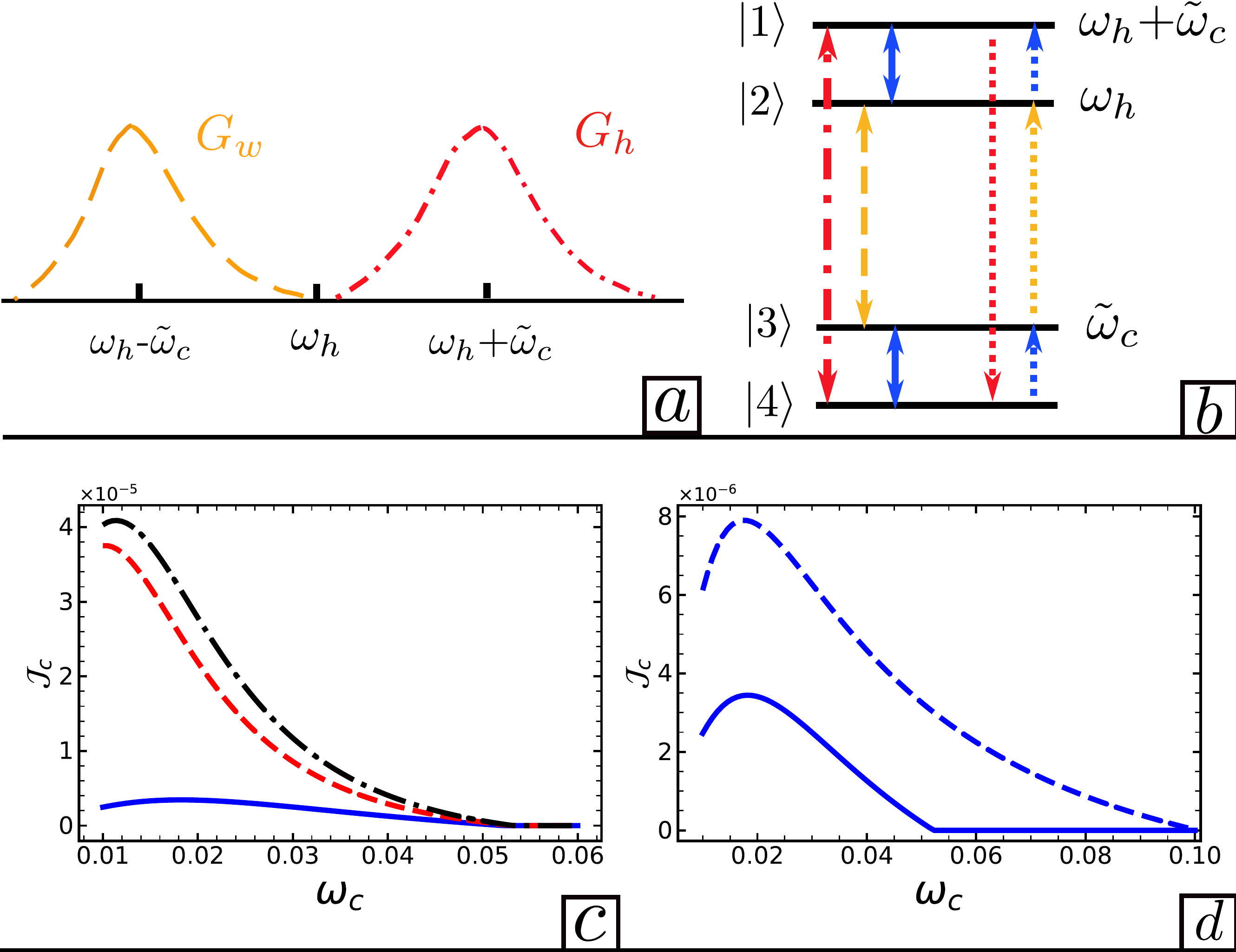}
		\caption{(Color online) (a) Shows the spectrally separated work (orange dashed) and hot (red dot-dashed) baths spectra given by Eq.~(\ref{eq:BathFilter}). Panel (b) shows the dressed energy levels of the coupled TLS and the double-ended arrows describe the possible transitions induced by the thermal baths. The orange dashed, red dot-dashed and blue solid arrows illustrate the transitions induced by the work-like reservoir, hot and cold baths, respectively. There is only one possible cooling cycle (43214), that consists of four transitions and it is shown in (b) by dotted single-ended arrows. 
		  (c), (d) cooling power $\mathcal{J}_{c}$ is plotted as a function of the subsystem B frequency $\omega_{c}$.   In panel (c), solid, dashed and dash-dotted lines are for TLS, TLOS and OMS, respectively. In (d), solid and dashed lines are for the choice of spectrally separated work and hot baths spectra given in panel (a) and Fig.~\ref{fig:qRefLevels}(c), respectively, for TLS.}
		\label{fig:FSQAR}
	\end{figure} 
(ii) {\it{Coupled two-level oscillator system}}: the diagonalized Hamiltonian for this system excluding thermal baths is given by
\begin{equation}\label{eq:acHami}
\tilde{H}_{qo} = \frac{\omega_{h}}{2}\tilde{\sigma^{z}_{h}} + \omega_{c}\tilde{b}^{\dagger}\tilde{b}-\frac{g^2}{\omega_{c}},
\end{equation}
the transformed operators in terms of the bare ones read as
\begin{eqnarray}
\tilde{\sigma}^{-}_{h} &=& \hat{U}\hat{\sigma}_{h}^{-}\hat{U}^{\dagger} = \hat{\sigma}^{-}_{h} e^{-\beta(\hat{b}^{\dagger}-\hat{b})}\\ \nonumber
\tilde{b} &=& \hat{U}\hat{b}\hat{U}^{\dagger} = \hat{b}-\beta\hat{\sigma}^{+}_{h}\hat{\sigma}^{-}_{h},
\end{eqnarray}
where the unitary transformation is described in Eq.~(\ref{eq:unitary}) with $\beta=g/\omega_{c}$. The eigenenergies of this system are $E_{n_{h}, m_{c}}=n_{h}\omega_{h}/2+m_{c}\omega_{{c}}-g^2/\omega_{c}$, where $n_{a}=0,1$ corresponds to ground and excited state of the two level system (TLS), respectively. Furthermore, $m_{c}$ is the cavity photon number and the cavity Fock state is displaced by a factor of $\beta$.

(iii) {\it{optomechanical system}}: The diagonalized Hamiltonian of an isolated optomechanical system consisting of a cavity and a mechanical mode is given by
\begin{equation}\label{eq:omHami}
\tilde{H}_{om} = \omega_{h}\tilde{a}^{\dagger}\tilde{a} + \omega_{c}\tilde{b}^{\dagger}\tilde{b}-\frac{g^2}{\omega_{c}}(\tilde{a}^{\dagger}\tilde{a})^2,
\end{equation}
where the dressed operators in terms of bare operators given by
\begin{eqnarray}
\tilde{a} &=& \hat{U}\hat{a}\hat{U}^{\dagger} = \hat{a} e^{-\beta(\hat{b}^{\dagger}-\hat{b})},\\ \nonumber
\tilde{b} &=& \hat{U}\hat{b}\hat{U}^{\dagger} = \hat{b}-\beta\hat{a}^{\dagger}\hat{a}.
\end{eqnarray}
The eigenenergies for the dressed optomechanial Hamiltonian are $E_{n_{h}, m_{c}}=n_{h}\omega_{h}+m_{c}\omega_{{c}}-ng^2/\omega_{c}$, where $n_{h}$ is the number of photons in the cavity and $m_{c}$ is phonon number.

\section{\label{AppendixB} Heat Currents for coupled qubits}
\label{app:dynamics}
The equations of motions to calculate the steady-state heat currents in case of coupled TLS for the choice of bath spectra shown in Fig.~\ref{fig:qRefLevels}(c) are determined from , Eq.~(\ref{eq:master}), and given by
\begin{eqnarray}\label{eq:dynamics}
\frac{d}{dt}\langle\tilde{\sigma}^{+}_{h}\tilde{\sigma}^{-}_{h}\rangle &=& -c^2\Gamma_{h}(1+e^{-\frac{\omega_{h}}{T_{h}}})\langle\tilde{\sigma}^{+}_{h}\tilde{\sigma}^{-}_{h}\rangle + c^2\Gamma_{h} e^{-\frac{\omega_{h}}{T_{h}}}\nonumber \\ &+& s^2 \Gamma_{w}(\langle\tilde{\sigma}^{+}_{c}\tilde{\sigma}^{-}_{c}\rangle - \langle\tilde{\sigma}^{+}_{h}\tilde{\sigma}^{-}_{h}\rangle),\\
\frac{d}{dt}\langle\tilde{\sigma}^{+}_{c}\tilde{\sigma}^{-}_{c}\rangle &=& -c^2\Gamma_{c}(1+e^{-\frac{\Omega}{T_{c}}})\langle\tilde{\sigma}^{+}_{c}\tilde{\sigma}^{-}_{c}\rangle + c^2\Gamma_{c} e^{-\frac{\Omega}{T_{c}}}\nonumber \\ &+& s^2\Gamma_{w}(\langle\tilde{\sigma}^{+}_{h}\tilde{\sigma}^{-}_{h}\rangle - \langle\tilde{\sigma}^{+}_{c}\tilde{\sigma}^{-}_{c}\rangle),
\end{eqnarray}
where we have considered $T_{w}\to \infty$, consequently, higher order correlations such as $\langle\tilde{\sigma}^{+}_{h}\tilde{\sigma}^{-}_{h}\tilde{\sigma}^{+}_{c}\tilde{\sigma}^{-}_{c}\rangle$ vanish. In addition, $c=$ cos$\theta$, $s=$ sin$\theta$, and
	\begin{eqnarray}\label{eq:gamma}
	\Gamma_{\alpha} = \omega_{\alpha}\kappa_{\alpha}\bar{n}(\omega_{\alpha}), 
	\end{eqnarray} 
The steady-state heat currents in the limint $T_{w}\to \infty$ are described as
\begin{equation}\label{eq:appheat}
\mathcal{J}_{\alpha} = \omega_{\alpha} K, 
\end{equation}
and
\begin{widetext}
\begin{eqnarray}
K &=& \frac{c^2s^2 \Gamma_{w}\Gamma_{h}\Gamma_{c}(e^{\frac{\tilde{\omega}_{c}}{T_{c}}}-e^{\frac{\tilde{\omega}_{c}}{T_{h}}})}{c^2 \Gamma_{w}\Gamma_{c}(1+e^{\frac{\tilde{\omega}_{c}}{T_{c}}})(1+e^{\frac{\omega_{h}}{T_{h}}}) + s^2\Gamma_{w}[e^{\frac{\omega_{h}}{T_{h}}}\Gamma_{c} +e^{\frac{\tilde{\omega}_{c}}{T_{c}}}(\Gamma_{w}+e^{\frac{\omega_{h}}{T_{h}}}(\Gamma_{w}+\Gamma_{c}))]}.
\end{eqnarray}
\end{widetext}

\section{\label{AppendixC} Single cooling cycle QAR}
\label{app:FSQAR}

Here we investigate the working of our QAR for the selection of different spectra of hot and work reservoirs than presented in the main text. We consider spectrally filtered spectra shown in Fig.~\ref{fig:FSQAR}(a), and the cold bath has OSD. For the case of TLS, under the selection of these bath spectra, only one energy cycle (43214) and its inverse are possible. The (43214) cycle consists of four transitions and if parameters are selected within the cooling window, then it is responsible for cooling. 
	\begin{figure}[t]
		\centering
		\includegraphics[width=0.45\textwidth]{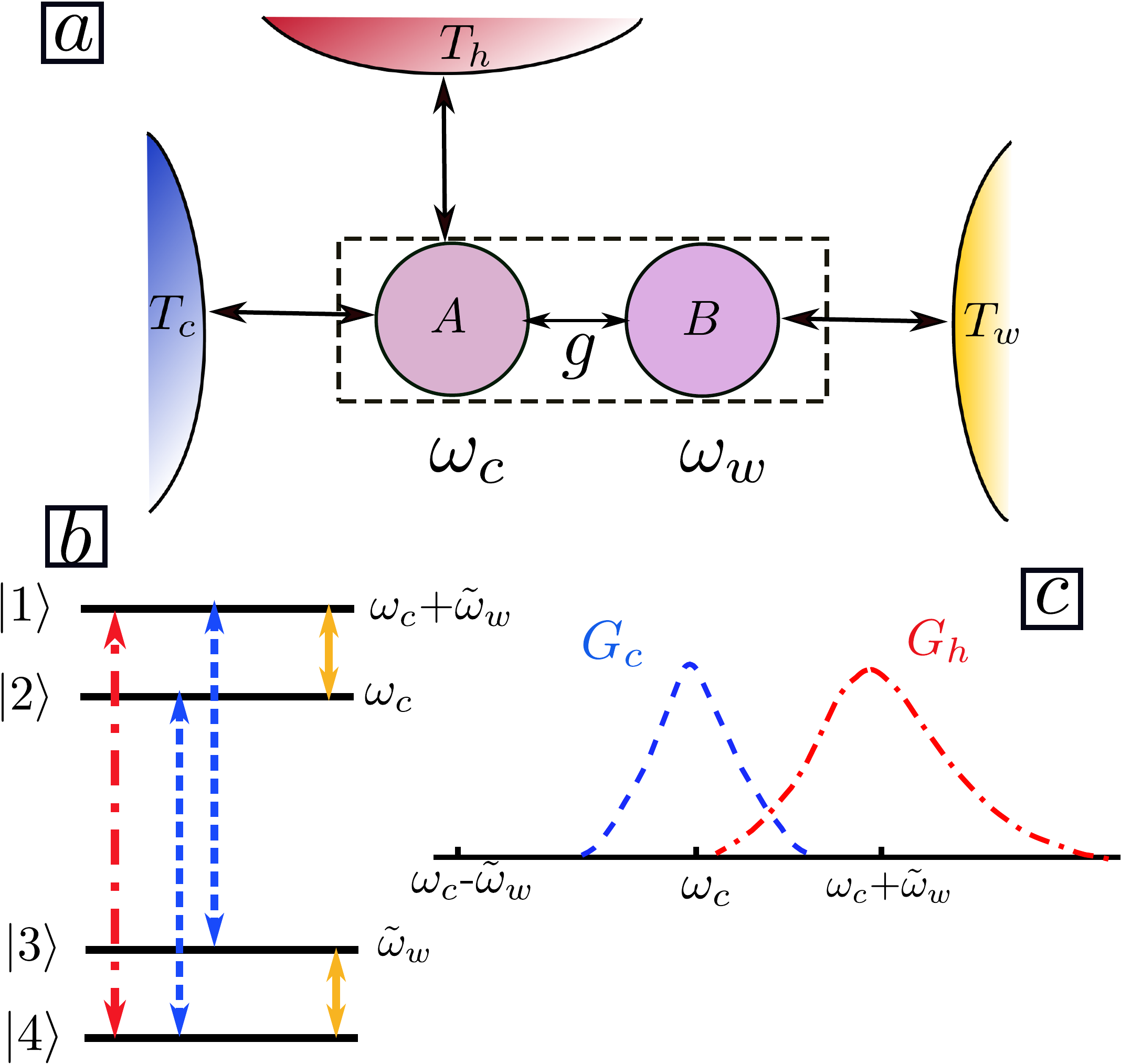}
		\caption{(Color online)  Schematic illustration of the QAR model, here work and cold baths are swapped compared to the model presented in Fig.~\ref{fig:jmodel}, and subsystem A (B) has a frequency $\omega_{c}$ ($\omega_{w}$). Panel (b) shows the dressed energy levels of the coupled TLS and the double-ended arrows describe the possible transitions induced by the thermal baths. The solid, dashed and dot-dashed arrows illustrate the transitions induced by the work-like reservoir, cold and hot baths, respectively. (c) shows the spectrally separated cold (dashed) and hot (dot-dashed) baths spectra given by Eq.~(\ref{eq:BathFilter})}
		\label{fig:wcEff}
	\end{figure} 
For the same baths spectra, similar four transitions energy cycles can also be identified for TLOS and OMS which lead to cooling.  To show it explicitly, we plot the cooling power $\mathcal{J}_{c}$ as a function of the subsystem B frequency $\omega_{c}$ for TLS, TLOS, and OMS in Fig.~\ref{fig:FSQAR}(c). The higher dimensional OMS outperforms the lower dimensional TLOS and TLS models. For comparison between the performance of three (Fig.~\ref{fig:qRefLevels}(a)) and four transitions (Fig.~\ref{fig:FSQAR}(b)) energy cycles models of TLS, we plot the cooling power $\mathcal{J}_{c}$ as a function of the subsystem B frequency $\omega_{c}$, shown in Fig.~\ref{fig:FSQAR}(d). The model based on three transitions energy cycles presented in the main text outperforms the four transitions energy cycles model. Because  the model presented Fig.~\ref{fig:qRefLevels}(a) can be considered as two three-level QARs working together, which can outperform a single four-level QAR. A four-level QAR, similar to the system presented in Fig.~\ref{fig:FSQAR}(b) has been proposed in the Ref.~\cite{PhysRevE.92.012136}, without giving a specific physical implementation. Our study shows that this can be implemented via optomechanical-like coupling between the coupled two two-level systems or two harmonic oscillators or a two-level system and a harmonic oscillator.   
\section{\label{AppendixD} Efficiency at maximum power and cold bath spectrum}
\label{app:wcEff}
	\begin{figure}[t]
	\hspace*{-0.50cm}
		\centering
		\includegraphics[width=0.50\textwidth]{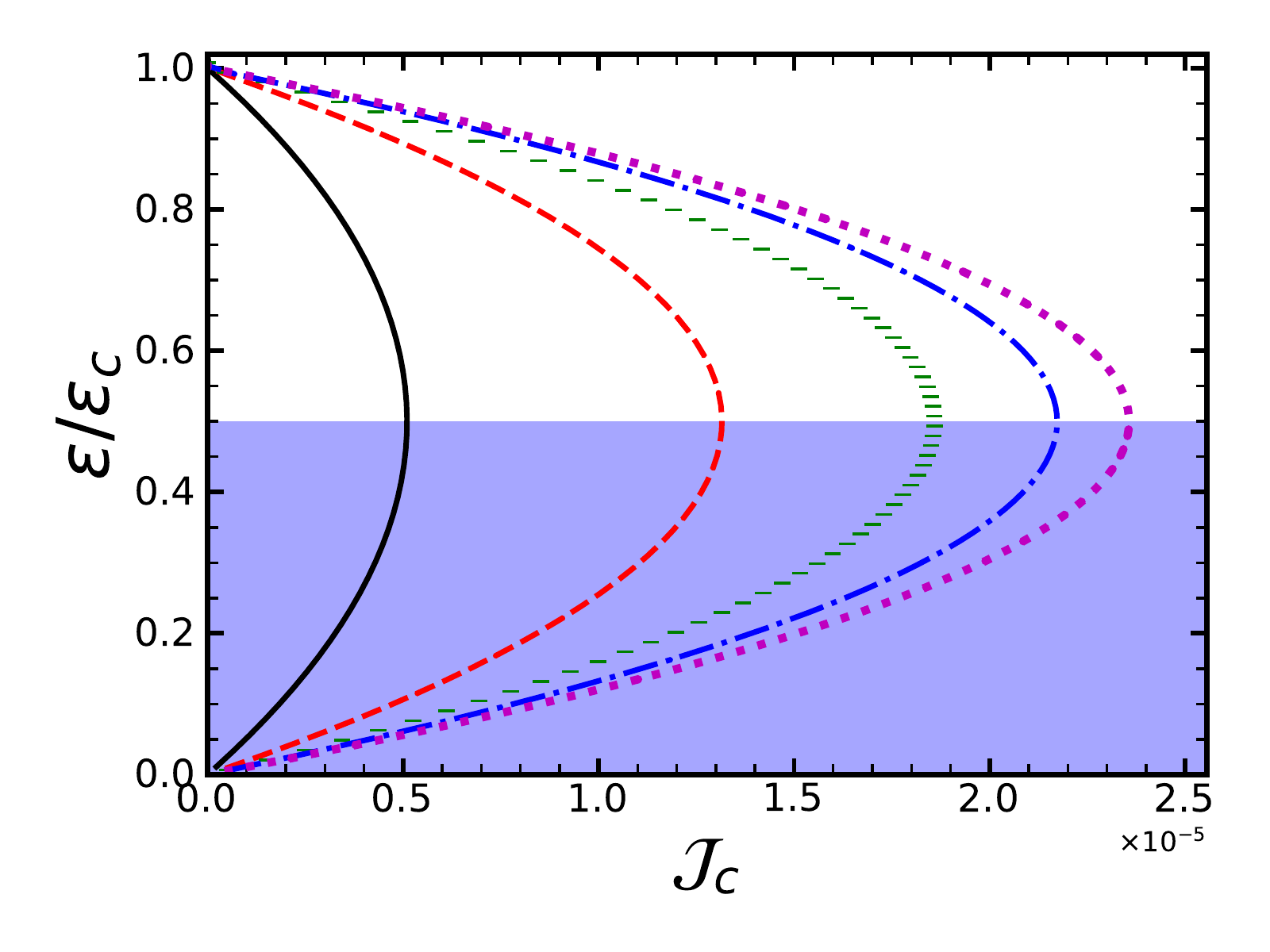}
		\caption{(Color online) Parametric plot of cooling power $\mathcal{J}_{c}$ vs. normalized coefficient of performance (COP) $\epsilon/\epsilon_{c}$ for different values of the coupling strength $g$. Solid, dashed, horizontal dashed, dash-dotted and dotted lines are for $g={0.025, 0.05, 0.075, 0.1, and ~0.125}$, respectively. Rest of the parameters are $T_{w}=10$,  $T_{h}=6$, $T_{c}=5$, $\omega_{w}=0.1$.
		The shaded area shows the efficiency at maximum power $\epsilon_{*}$ bound for one-dimensional bosonic reservoirs given in Eq.~(\ref{eq:ebound}).}
		\label{fig:COPMPApp}
	\end{figure} 
In the main text, we argue that the increase in the efficiency at maximum power $\epsilon_{*}$ in our model is because the dimensionality of the cold bath spectrum effectively changes, which bounds the $\epsilon_{*}$, as given in Eq.~(\ref{eq:ebound}). To emphasize this, we consider a case shown in Fig.~\ref{fig:wcEff}, in which the cold bath dimensionality does not change effectively. Note that, we have swapped the cold and work reservoir compared to the model in Fig.~\ref{fig:jmodel}, in addition, now subsystem A (B) frequency is $\omega_{c}$ ($\omega_{w}$). To avoid confusion, we re-write the master equation given in Eq.~(\ref{eq:master}) for this case,
\begin{eqnarray}
	\label{eq:L_LAp}
\hat{ \mathcal{L}}_{k} &=& G_{k}(\omega_{c})c^{2}\hat{\mathcal{D}}[\tilde{a}]
	+ G_{k}(-\omega_{c})c^{2}\hat{\mathcal{D}}[\tilde{a}^{\dagger}] \\ \nonumber&+& G_{k}(\omega_{-})s^{2}\hat{\mathcal{D}}[\tilde{a}\tilde{b}^{\dagger}]
	+ G_{k}(-\omega_{-})s^{2}\hat{\mathcal{D}}[\tilde{a}^{\dagger}\tilde{b}]
	\\ \nonumber &+& G_{k}(\omega_{h})s^{2}\hat{\mathcal{D}}[\tilde{a}\tilde{b}]
	+ G_{k}(-\omega_{h})s^{2}\hat{\mathcal{D}}[\tilde{a}^{\dagger}\tilde{b}^{\dagger}],
	\\
	\hat{ \mathcal{L}}_{w}&=& G_{w}(\tilde{\omega}_{w})c^{2}\hat{\mathcal{D}}[\tilde{b}]
	+ G_{w}(-\tilde{\omega}_{w})c^{2}\hat{\mathcal{D}}[\tilde{b}^{\dagger}], \label{eq:L_MAp}
\end{eqnarray}
where, $s=2g/\omega_{w}$, $c=\omega_{w}/\tilde{\omega}_{w}$, $\omega_{-}=\omega_{c}-\omega_{h}$, $\tilde{\omega}_{w}=\sqrt{\omega_{w}+4g^2}$, and $k=h, c$ for hot and cold baths, respectively. The possible transitions induced by the three baths are shown in Fig.~\ref{fig:wcEff}(b), under the selection of the hot and cold baths spectrums presented in Fig.~\ref{fig:wcEff}(c). Note that the dimensonality of
cold bath spectrum $G_{c}(\omega_{c})$ does not change by the presence of $s$ and $c$ terms in the Eq.~(\ref{eq:L_LAp}). In this case, the bound given in Eq.~(\ref{eq:ebound}) should hold, and this is verified in Fig.~\ref{fig:COPMPApp}, where we plot cooling power $\mathcal{J}_{c}$ as a function of the normalized coefficient of performance $\epsilon/\epsilon_{c}$ for different values of coupling strengths. Unlike the case presented in Fig.~\ref{fig:COPMP}, here the coupling strength $g$ does not change $\omega_{c}$, consequently does not affect the $\epsilon_{*}$.

%

\begin{thebibliography}{0}%
\makeatletter
\providecommand \@ifxundefined [1]{%
 \@ifx{#1\undefined}
}%
\providecommand \@ifnum [1]{%
 \ifnum #1\expandafter \@firstoftwo
 \else \expandafter \@secondoftwo
 \fi
}%
\providecommand \@ifx [1]{%
 \ifx #1\expandafter \@firstoftwo
 \else \expandafter \@secondoftwo
 \fi
}%
\providecommand \natexlab [1]{#1}%
\providecommand \enquote  [1]{``#1''}%
\providecommand \bibnamefont  [1]{#1}%
\providecommand \bibfnamefont [1]{#1}%
\providecommand \citenamefont [1]{#1}%
\providecommand \href@noop [0]{\@secondoftwo}%
\providecommand \href [0]{\begingroup \@sanitize@url \@href}%
\providecommand \@href[1]{\@@startlink{#1}\@@href}%
\providecommand \@@href[1]{\endgroup#1\@@endlink}%
\providecommand \@sanitize@url [0]{\catcode `\\12\catcode `\$12\catcode
  `\&12\catcode `\#12\catcode `\^12\catcode `\_12\catcode `\%12\relax}%
\providecommand \@@startlink[1]{}%
\providecommand \@@endlink[0]{}%
\providecommand \url  [0]{\begingroup\@sanitize@url \@url }%
\providecommand \@url [1]{\endgroup\@href {#1}{\urlprefix }}%
\providecommand \urlprefix  [0]{URL }%
\providecommand \Eprint [0]{\href }%
\providecommand \doibase [0]{http://dx.doi.org/}%
\providecommand \selectlanguage [0]{\@gobble}%
\providecommand \bibinfo  [0]{\@secondoftwo}%
\providecommand \bibfield  [0]{\@secondoftwo}%
\providecommand \translation [1]{[#1]}%
\providecommand \BibitemOpen [0]{}%
\providecommand \bibitemStop [0]{}%
\providecommand \bibitemNoStop [0]{.\EOS\space}%
\providecommand \EOS [0]{\spacefactor3000\relax}%
\providecommand \BibitemShut  [1]{\csname bibitem#1\endcsname}%
\let\auto@bib@innerbib\@empty
\end{thebibliography}%


\begin{thebibliography}{55}%
\makeatletter
\providecommand \@ifxundefined [1]{%
 \@ifx{#1\undefined}
}%
\providecommand \@ifnum [1]{%
 \ifnum #1\expandafter \@firstoftwo
 \else \expandafter \@secondoftwo
 \fi
}%
\providecommand \@ifx [1]{%
 \ifx #1\expandafter \@firstoftwo
 \else \expandafter \@secondoftwo
 \fi
}%
\providecommand \natexlab [1]{#1}%
\providecommand \enquote  [1]{``#1''}%
\providecommand \bibnamefont  [1]{#1}%
\providecommand \bibfnamefont [1]{#1}%
\providecommand \citenamefont [1]{#1}%
\providecommand \href@noop [0]{\@secondoftwo}%
\providecommand \href [0]{\begingroup \@sanitize@url \@href}%
\providecommand \@href[1]{\@@startlink{#1}\@@href}%
\providecommand \@@href[1]{\endgroup#1\@@endlink}%
\providecommand \@sanitize@url [0]{\catcode `\\12\catcode `\$12\catcode
  `\&12\catcode `\#12\catcode `\^12\catcode `\_12\catcode `\%12\relax}%
\providecommand \@@startlink[1]{}%
\providecommand \@@endlink[0]{}%
\providecommand \url  [0]{\begingroup\@sanitize@url \@url }%
\providecommand \@url [1]{\endgroup\@href {#1}{\urlprefix }}%
\providecommand \urlprefix  [0]{URL }%
\providecommand \Eprint [0]{\href }%
\providecommand \doibase [0]{http://dx.doi.org/}%
\providecommand \selectlanguage [0]{\@gobble}%
\providecommand \bibinfo  [0]{\@secondoftwo}%
\providecommand \bibfield  [0]{\@secondoftwo}%
\providecommand \translation [1]{[#1]}%
\providecommand \BibitemOpen [0]{}%
\providecommand \bibitemStop [0]{}%
\providecommand \bibitemNoStop [0]{.\EOS\space}%
\providecommand \EOS [0]{\spacefactor3000\relax}%
\providecommand \BibitemShut  [1]{\csname bibitem#1\endcsname}%
\let\auto@bib@innerbib\@empty
\bibitem [{\citenamefont {Kosloff}\ and\ \citenamefont
  {Levy}(2014)}]{doi:10.1146/annurev-physchem-040513-103724}%
  \BibitemOpen
  \bibfield  {author} {\bibinfo {author} {\bibfnamefont {Ronnie}\ \bibnamefont
  {Kosloff}}\ and\ \bibinfo {author} {\bibfnamefont {Amikam}\ \bibnamefont
  {Levy}},\ }\bibfield  {title} {\enquote {\bibinfo {title} {Quantum heat
  engines and refrigerators: Continuous devices},}\ }\href {\doibase
  10.1146/annurev-physchem-040513-103724} {\bibfield  {journal} {\bibinfo
  {journal} {Annu. Rev. Phys. Chem.}\ }\textbf {\bibinfo {volume} {65}},\
  \bibinfo {pages} {365--393} (\bibinfo {year} {2014})}\BibitemShut {NoStop}%
\bibitem [{\citenamefont {Ro{\ss}nagel}\ \emph {et~al.}(2016)\citenamefont
  {Ro{\ss}nagel}, \citenamefont {Dawkins}, \citenamefont {Tolazzi},
  \citenamefont {Abah}, \citenamefont {Lutz}, \citenamefont {Schmidt-Kaler},\
  and\ \citenamefont {Singer}}]{Rossnagel325}%
  \BibitemOpen
  \bibfield  {author} {\bibinfo {author} {\bibfnamefont {Johannes}\
  \bibnamefont {Ro{\ss}nagel}}, \bibinfo {author} {\bibfnamefont {Samuel~T.}\
  \bibnamefont {Dawkins}}, \bibinfo {author} {\bibfnamefont {Karl~N.}\
  \bibnamefont {Tolazzi}}, \bibinfo {author} {\bibfnamefont {Obinna}\
  \bibnamefont {Abah}}, \bibinfo {author} {\bibfnamefont {Eric}\ \bibnamefont
  {Lutz}}, \bibinfo {author} {\bibfnamefont {Ferdinand}\ \bibnamefont
  {Schmidt-Kaler}}, \ and\ \bibinfo {author} {\bibfnamefont {Kilian}\
  \bibnamefont {Singer}},\ }\bibfield  {title} {\enquote {\bibinfo {title} {A
  single-atom heat engine},}\ }\href {\doibase 10.1126/science.aad6320}
  {\bibfield  {journal} {\bibinfo  {journal} {Science}\ }\textbf {\bibinfo
  {volume} {352}},\ \bibinfo {pages} {325--329} (\bibinfo {year}
  {2016})}\BibitemShut {NoStop}%
\bibitem [{\citenamefont {Hardal}\ \emph {et~al.}(2017)\citenamefont {Hardal},
  \citenamefont {Aslan}, \citenamefont {Wilson},\ and\ \citenamefont
  {M\"{u}stecapl{\i}\u{g}lu}}]{PhysRevE.96.062120}%
  \BibitemOpen
  \bibfield  {author} {\bibinfo {author} {\bibfnamefont {Ali \"{U}.~C.}\
  \bibnamefont {Hardal}}, \bibinfo {author} {\bibfnamefont {Nur}\ \bibnamefont
  {Aslan}}, \bibinfo {author} {\bibfnamefont {C.~M.}\ \bibnamefont {Wilson}}, \
  and\ \bibinfo {author} {\bibfnamefont {\"{O}zg\"{u}r~E.}\ \bibnamefont
  {M\"{u}stecapl{\i}\u{g}lu}},\ }\bibfield  {title} {\enquote {\bibinfo {title}
  {Quantum heat engine with coupled superconducting resonators},}\ }\href
  {\doibase 10.1103/PhysRevE.96.062120} {\bibfield  {journal} {\bibinfo
  {journal} {Phys. Rev. E}\ }\textbf {\bibinfo {volume} {96}},\ \bibinfo
  {pages} {062120} (\bibinfo {year} {2017})}\BibitemShut {NoStop}%
\bibitem [{\citenamefont {Karimi}\ and\ \citenamefont
  {Pekola}(2016)}]{PhysRevB.94.184503}%
  \BibitemOpen
  \bibfield  {author} {\bibinfo {author} {\bibfnamefont {B.}~\bibnamefont
  {Karimi}}\ and\ \bibinfo {author} {\bibfnamefont {J.~P.}\ \bibnamefont
  {Pekola}},\ }\bibfield  {title} {\enquote {\bibinfo {title} {Otto
  refrigerator based on a superconducting qubit: Classical and quantum
  performance},}\ }\href {\doibase 10.1103/PhysRevB.94.184503} {\bibfield
  {journal} {\bibinfo  {journal} {Phys. Rev. B}\ }\textbf {\bibinfo {volume}
  {94}},\ \bibinfo {pages} {184503} (\bibinfo {year} {2016})}\BibitemShut
  {NoStop}%
\bibitem [{\citenamefont {Manikandan}\ \emph {et~al.}(2019)\citenamefont
  {Manikandan}, \citenamefont {Giazotto},\ and\ \citenamefont
  {Jordan}}]{PhysRevApplied.11.054034}%
  \BibitemOpen
  \bibfield  {author} {\bibinfo {author} {\bibfnamefont {Sreenath~K.}\
  \bibnamefont {Manikandan}}, \bibinfo {author} {\bibfnamefont {Francesco}\
  \bibnamefont {Giazotto}}, \ and\ \bibinfo {author} {\bibfnamefont
  {Andrew~N.}\ \bibnamefont {Jordan}},\ }\bibfield  {title} {\enquote {\bibinfo
  {title} {Superconducting quantum refrigerator: Breaking and rejoining cooper
  pairs with magnetic field cycles},}\ }\href {\doibase
  10.1103/PhysRevApplied.11.054034} {\bibfield  {journal} {\bibinfo  {journal}
  {Phys. Rev. Appl.}\ }\textbf {\bibinfo {volume} {11}},\ \bibinfo {pages}
  {054034} (\bibinfo {year} {2019})}\BibitemShut {NoStop}%
\bibitem [{\citenamefont
  {Mitchison}(2019)}]{doi:10.1080/00107514.2019.1631555}%
  \BibitemOpen
  \bibfield  {author} {\bibinfo {author} {\bibfnamefont {Mark~T.}\ \bibnamefont
  {Mitchison}},\ }\bibfield  {title} {\enquote {\bibinfo {title} {Quantum
  thermal absorption machines: refrigerators, engines and clocks},}\ }\href
  {\doibase 10.1080/00107514.2019.1631555} {\bibfield  {journal} {\bibinfo
  {journal} {Contemp. Phys.}\ }\textbf {\bibinfo {volume} {60}},\ \bibinfo
  {pages} {164--187} (\bibinfo {year} {2019})}\BibitemShut {NoStop}%
\bibitem [{\citenamefont {Palao}\ \emph {et~al.}(2001)\citenamefont {Palao},
  \citenamefont {Kosloff},\ and\ \citenamefont {Gordon}}]{PhysRevE.64.056130}%
  \BibitemOpen
  \bibfield  {author} {\bibinfo {author} {\bibfnamefont {Jos\'e~P.}\
  \bibnamefont {Palao}}, \bibinfo {author} {\bibfnamefont {Ronnie}\
  \bibnamefont {Kosloff}}, \ and\ \bibinfo {author} {\bibfnamefont
  {Jeffrey~M.}\ \bibnamefont {Gordon}},\ }\bibfield  {title} {\enquote
  {\bibinfo {title} {Quantum thermodynamic cooling cycle},}\ }\href {\doibase
  10.1103/PhysRevE.64.056130} {\bibfield  {journal} {\bibinfo  {journal} {Phys.
  Rev. E}\ }\textbf {\bibinfo {volume} {64}},\ \bibinfo {pages} {056130}
  (\bibinfo {year} {2001})}\BibitemShut {NoStop}%
\bibitem [{\citenamefont {Linden}\ \emph {et~al.}(2010)\citenamefont {Linden},
  \citenamefont {Popescu},\ and\ \citenamefont
  {Skrzypczyk}}]{PhysRevLett.105.130401}%
  \BibitemOpen
  \bibfield  {author} {\bibinfo {author} {\bibfnamefont {Noah}\ \bibnamefont
  {Linden}}, \bibinfo {author} {\bibfnamefont {Sandu}\ \bibnamefont {Popescu}},
  \ and\ \bibinfo {author} {\bibfnamefont {Paul}\ \bibnamefont {Skrzypczyk}},\
  }\bibfield  {title} {\enquote {\bibinfo {title} {How small can thermal
  machines be the smallest possible refrigerator},}\ }\href {\doibase
  10.1103/PhysRevLett.105.130401} {\bibfield  {journal} {\bibinfo  {journal}
  {Phys. Rev. Lett.}\ }\textbf {\bibinfo {volume} {105}},\ \bibinfo {pages}
  {130401} (\bibinfo {year} {2010})}\BibitemShut {NoStop}%
\bibitem [{\citenamefont {Levy}\ and\ \citenamefont
  {Kosloff}(2012)}]{PhysRevLett.108.070604}%
  \BibitemOpen
  \bibfield  {author} {\bibinfo {author} {\bibfnamefont {Amikam}\ \bibnamefont
  {Levy}}\ and\ \bibinfo {author} {\bibfnamefont {Ronnie}\ \bibnamefont
  {Kosloff}},\ }\bibfield  {title} {\enquote {\bibinfo {title} {Quantum
  absorption refrigerator},}\ }\href {\doibase 10.1103/PhysRevLett.108.070604}
  {\bibfield  {journal} {\bibinfo  {journal} {Phys. Rev. Lett.}\ }\textbf
  {\bibinfo {volume} {108}},\ \bibinfo {pages} {070604} (\bibinfo {year}
  {2012})}\BibitemShut {NoStop}%
\bibitem [{\citenamefont {Levy}\ \emph {et~al.}(2012)\citenamefont {Levy},
  \citenamefont {Alicki},\ and\ \citenamefont {Kosloff}}]{PhysRevE.85.061126}%
  \BibitemOpen
  \bibfield  {author} {\bibinfo {author} {\bibfnamefont {Amikam}\ \bibnamefont
  {Levy}}, \bibinfo {author} {\bibfnamefont {Robert}\ \bibnamefont {Alicki}}, \
  and\ \bibinfo {author} {\bibfnamefont {Ronnie}\ \bibnamefont {Kosloff}},\
  }\bibfield  {title} {\enquote {\bibinfo {title} {Quantum refrigerators and
  the third law of thermodynamics},}\ }\href {\doibase
  10.1103/PhysRevE.85.061126} {\bibfield  {journal} {\bibinfo  {journal} {Phys.
  Rev. E}\ }\textbf {\bibinfo {volume} {85}},\ \bibinfo {pages} {061126}
  (\bibinfo {year} {2012})}\BibitemShut {NoStop}%
\bibitem [{\citenamefont {Correa}\ \emph {et~al.}(2013)\citenamefont {Correa},
  \citenamefont {Palao}, \citenamefont {Adesso},\ and\ \citenamefont
  {Alonso}}]{PhysRevE.87.042131}%
  \BibitemOpen
  \bibfield  {author} {\bibinfo {author} {\bibfnamefont {Luis~A.}\ \bibnamefont
  {Correa}}, \bibinfo {author} {\bibfnamefont {Jos\'e~P.}\ \bibnamefont
  {Palao}}, \bibinfo {author} {\bibfnamefont {Gerardo}\ \bibnamefont {Adesso}},
  \ and\ \bibinfo {author} {\bibfnamefont {Daniel}\ \bibnamefont {Alonso}},\
  }\bibfield  {title} {\enquote {\bibinfo {title} {Performance bound for
  quantum absorption refrigerators},}\ }\href {\doibase
  10.1103/PhysRevE.87.042131} {\bibfield  {journal} {\bibinfo  {journal} {Phys.
  Rev. E}\ }\textbf {\bibinfo {volume} {87}},\ \bibinfo {pages} {042131}
  (\bibinfo {year} {2013})}\BibitemShut {NoStop}%
\bibitem [{\citenamefont {Venturelli}\ \emph {et~al.}(2013)\citenamefont
  {Venturelli}, \citenamefont {Fazio},\ and\ \citenamefont
  {Giovannetti}}]{PhysRevLett.110.256801}%
  \BibitemOpen
  \bibfield  {author} {\bibinfo {author} {\bibfnamefont {Davide}\ \bibnamefont
  {Venturelli}}, \bibinfo {author} {\bibfnamefont {Rosario}\ \bibnamefont
  {Fazio}}, \ and\ \bibinfo {author} {\bibfnamefont {Vittorio}\ \bibnamefont
  {Giovannetti}},\ }\bibfield  {title} {\enquote {\bibinfo {title} {Minimal
  self-contained quantum refrigeration machine based on four quantum dots},}\
  }\href {\doibase 10.1103/PhysRevLett.110.256801} {\bibfield  {journal}
  {\bibinfo  {journal} {Phys. Rev. Lett.}\ }\textbf {\bibinfo {volume} {110}},\
  \bibinfo {pages} {256801} (\bibinfo {year} {2013})}\BibitemShut {NoStop}%
\bibitem [{\citenamefont {Correa}(2014)}]{PhysRevE.89.042128}%
  \BibitemOpen
  \bibfield  {author} {\bibinfo {author} {\bibfnamefont {Luis~A.}\ \bibnamefont
  {Correa}},\ }\bibfield  {title} {\enquote {\bibinfo {title} {Multistage
  quantum absorption heat pumps},}\ }\href {\doibase
  10.1103/PhysRevE.89.042128} {\bibfield  {journal} {\bibinfo  {journal} {Phys.
  Rev. E}\ }\textbf {\bibinfo {volume} {89}},\ \bibinfo {pages} {042128}
  (\bibinfo {year} {2014})}\BibitemShut {NoStop}%
\bibitem [{\citenamefont {Leggio}\ \emph {et~al.}(2015)\citenamefont {Leggio},
  \citenamefont {Bellomo},\ and\ \citenamefont {Antezza}}]{PhysRevA.91.012117}%
  \BibitemOpen
  \bibfield  {author} {\bibinfo {author} {\bibfnamefont {Bruno}\ \bibnamefont
  {Leggio}}, \bibinfo {author} {\bibfnamefont {Bruno}\ \bibnamefont {Bellomo}},
  \ and\ \bibinfo {author} {\bibfnamefont {Mauro}\ \bibnamefont {Antezza}},\
  }\bibfield  {title} {\enquote {\bibinfo {title} {Quantum thermal machines
  with single nonequilibrium environments},}\ }\href {\doibase
  10.1103/PhysRevA.91.012117} {\bibfield  {journal} {\bibinfo  {journal} {Phys.
  Rev. A}\ }\textbf {\bibinfo {volume} {91}},\ \bibinfo {pages} {012117}
  (\bibinfo {year} {2015})}\BibitemShut {NoStop}%
\bibitem [{\citenamefont {Silva}\ \emph {et~al.}(2015)\citenamefont {Silva},
  \citenamefont {Skrzypczyk},\ and\ \citenamefont
  {Brunner}}]{PhysRevE.92.012136}%
  \BibitemOpen
  \bibfield  {author} {\bibinfo {author} {\bibfnamefont {Ralph}\ \bibnamefont
  {Silva}}, \bibinfo {author} {\bibfnamefont {Paul}\ \bibnamefont
  {Skrzypczyk}}, \ and\ \bibinfo {author} {\bibfnamefont {Nicolas}\
  \bibnamefont {Brunner}},\ }\bibfield  {title} {\enquote {\bibinfo {title}
  {Small quantum absorption refrigerator with reversed couplings},}\ }\href
  {\doibase 10.1103/PhysRevE.92.012136} {\bibfield  {journal} {\bibinfo
  {journal} {Phys. Rev. E}\ }\textbf {\bibinfo {volume} {92}},\ \bibinfo
  {pages} {012136} (\bibinfo {year} {2015})}\BibitemShut {NoStop}%
\bibitem [{\citenamefont {Silva}\ \emph {et~al.}(2016)\citenamefont {Silva},
  \citenamefont {Manzano}, \citenamefont {Skrzypczyk},\ and\ \citenamefont
  {Brunner}}]{PhysRevE.94.032120}%
  \BibitemOpen
  \bibfield  {author} {\bibinfo {author} {\bibfnamefont {Ralph}\ \bibnamefont
  {Silva}}, \bibinfo {author} {\bibfnamefont {Gonzalo}\ \bibnamefont
  {Manzano}}, \bibinfo {author} {\bibfnamefont {Paul}\ \bibnamefont
  {Skrzypczyk}}, \ and\ \bibinfo {author} {\bibfnamefont {Nicolas}\
  \bibnamefont {Brunner}},\ }\bibfield  {title} {\enquote {\bibinfo {title}
  {Performance of autonomous quantum thermal machines: Hilbert space dimension
  as a thermodynamical resource},}\ }\href {\doibase
  10.1103/PhysRevE.94.032120} {\bibfield  {journal} {\bibinfo  {journal} {Phys.
  Rev. E}\ }\textbf {\bibinfo {volume} {94}},\ \bibinfo {pages} {032120}
  (\bibinfo {year} {2016})}\BibitemShut {NoStop}%
\bibitem [{\citenamefont {Hofer}\ \emph {et~al.}(2016)\citenamefont {Hofer},
  \citenamefont {Perarnau-Llobet}, \citenamefont {Brask}, \citenamefont
  {Silva}, \citenamefont {Huber},\ and\ \citenamefont
  {Brunner}}]{PhysRevB.94.235420}%
  \BibitemOpen
  \bibfield  {author} {\bibinfo {author} {\bibfnamefont {Patrick~P.}\
  \bibnamefont {Hofer}}, \bibinfo {author} {\bibfnamefont {Mart\'{\i}}\
  \bibnamefont {Perarnau-Llobet}}, \bibinfo {author} {\bibfnamefont
  {Jonatan~Bohr}\ \bibnamefont {Brask}}, \bibinfo {author} {\bibfnamefont
  {Ralph}\ \bibnamefont {Silva}}, \bibinfo {author} {\bibfnamefont {Marcus}\
  \bibnamefont {Huber}}, \ and\ \bibinfo {author} {\bibfnamefont {Nicolas}\
  \bibnamefont {Brunner}},\ }\bibfield  {title} {\enquote {\bibinfo {title}
  {Autonomous quantum refrigerator in a circuit qed architecture based on a
  josephson junction},}\ }\href {\doibase 10.1103/PhysRevB.94.235420}
  {\bibfield  {journal} {\bibinfo  {journal} {Phys. Rev. B}\ }\textbf {\bibinfo
  {volume} {94}},\ \bibinfo {pages} {235420} (\bibinfo {year}
  {2016})}\BibitemShut {NoStop}%
\bibitem [{\citenamefont {Doyeux}\ \emph {et~al.}(2016)\citenamefont {Doyeux},
  \citenamefont {Leggio}, \citenamefont {Messina},\ and\ \citenamefont
  {Antezza}}]{PhysRevE.93.022134}%
  \BibitemOpen
  \bibfield  {author} {\bibinfo {author} {\bibfnamefont {Pierre}\ \bibnamefont
  {Doyeux}}, \bibinfo {author} {\bibfnamefont {Bruno}\ \bibnamefont {Leggio}},
  \bibinfo {author} {\bibfnamefont {Riccardo}\ \bibnamefont {Messina}}, \ and\
  \bibinfo {author} {\bibfnamefont {Mauro}\ \bibnamefont {Antezza}},\
  }\bibfield  {title} {\enquote {\bibinfo {title} {Quantum thermal machine
  acting on a many-body quantum system: Role of correlations in thermodynamic
  tasks},}\ }\href {\doibase 10.1103/PhysRevE.93.022134} {\bibfield  {journal}
  {\bibinfo  {journal} {Phys. Rev. E}\ }\textbf {\bibinfo {volume} {93}},\
  \bibinfo {pages} {022134} (\bibinfo {year} {2016})}\BibitemShut {NoStop}%
\bibitem [{\citenamefont {Roulet}\ \emph {et~al.}(2017)\citenamefont {Roulet},
  \citenamefont {Nimmrichter}, \citenamefont {Arrazola}, \citenamefont {Seah},\
  and\ \citenamefont {Scarani}}]{PhysRevE.95.062131}%
  \BibitemOpen
  \bibfield  {author} {\bibinfo {author} {\bibfnamefont {Alexandre}\
  \bibnamefont {Roulet}}, \bibinfo {author} {\bibfnamefont {Stefan}\
  \bibnamefont {Nimmrichter}}, \bibinfo {author} {\bibfnamefont {Juan~Miguel}\
  \bibnamefont {Arrazola}}, \bibinfo {author} {\bibfnamefont {Stella}\
  \bibnamefont {Seah}}, \ and\ \bibinfo {author} {\bibfnamefont {Valerio}\
  \bibnamefont {Scarani}},\ }\bibfield  {title} {\enquote {\bibinfo {title}
  {Autonomous rotor heat engine},}\ }\href {\doibase
  10.1103/PhysRevE.95.062131} {\bibfield  {journal} {\bibinfo  {journal} {Phys.
  Rev. E}\ }\textbf {\bibinfo {volume} {95}},\ \bibinfo {pages} {062131}
  (\bibinfo {year} {2017})}\BibitemShut {NoStop}%
\bibitem [{\citenamefont {Erdman}\ \emph {et~al.}(2018)\citenamefont {Erdman},
  \citenamefont {Bhandari}, \citenamefont {Fazio}, \citenamefont {Pekola},\
  and\ \citenamefont {Taddei}}]{PhysRevB.98.045433}%
  \BibitemOpen
  \bibfield  {author} {\bibinfo {author} {\bibfnamefont {Paolo~Andrea}\
  \bibnamefont {Erdman}}, \bibinfo {author} {\bibfnamefont {Bibek}\
  \bibnamefont {Bhandari}}, \bibinfo {author} {\bibfnamefont {Rosario}\
  \bibnamefont {Fazio}}, \bibinfo {author} {\bibfnamefont {Jukka~P.}\
  \bibnamefont {Pekola}}, \ and\ \bibinfo {author} {\bibfnamefont {Fabio}\
  \bibnamefont {Taddei}},\ }\bibfield  {title} {\enquote {\bibinfo {title}
  {Absorption refrigerators based on coulomb-coupled single-electron
  systems},}\ }\href {\doibase 10.1103/PhysRevB.98.045433} {\bibfield
  {journal} {\bibinfo  {journal} {Phys. Rev. B}\ }\textbf {\bibinfo {volume}
  {98}},\ \bibinfo {pages} {045433} (\bibinfo {year} {2018})}\BibitemShut
  {NoStop}%
\bibitem [{\citenamefont {H\"{a}rtle}\ \emph {et~al.}(2018)\citenamefont
  {H\"{a}rtle}, \citenamefont {Schinabeck}, \citenamefont {Kulkarni},
  \citenamefont {Gelbwaser-Klimovsky}, \citenamefont {Thoss},\ and\
  \citenamefont {Peskin}}]{PhysRevB.98.081404}%
  \BibitemOpen
  \bibfield  {author} {\bibinfo {author} {\bibfnamefont {R.}~\bibnamefont
  {H\"{a}rtle}}, \bibinfo {author} {\bibfnamefont {C.}~\bibnamefont
  {Schinabeck}}, \bibinfo {author} {\bibfnamefont {M.}~\bibnamefont
  {Kulkarni}}, \bibinfo {author} {\bibfnamefont {D.}~\bibnamefont
  {Gelbwaser-Klimovsky}}, \bibinfo {author} {\bibfnamefont {M.}~\bibnamefont
  {Thoss}}, \ and\ \bibinfo {author} {\bibfnamefont {U.}~\bibnamefont
  {Peskin}},\ }\bibfield  {title} {\enquote {\bibinfo {title} {Cooling by
  heating in nonequilibrium nanosystems},}\ }\href {\doibase
  10.1103/PhysRevB.98.081404} {\bibfield  {journal} {\bibinfo  {journal} {Phys.
  Rev. B}\ }\textbf {\bibinfo {volume} {98}},\ \bibinfo {pages} {081404}
  (\bibinfo {year} {2018})}\BibitemShut {NoStop}%
\bibitem [{\citenamefont {Segal}(2018)}]{PhysRevE.97.052145}%
  \BibitemOpen
  \bibfield  {author} {\bibinfo {author} {\bibfnamefont {Dvira}\ \bibnamefont
  {Segal}},\ }\bibfield  {title} {\enquote {\bibinfo {title} {Current
  fluctuations in quantum absorption refrigerators},}\ }\href {\doibase
  10.1103/PhysRevE.97.052145} {\bibfield  {journal} {\bibinfo  {journal} {Phys.
  Rev. E}\ }\textbf {\bibinfo {volume} {97}},\ \bibinfo {pages} {052145}
  (\bibinfo {year} {2018})}\BibitemShut {NoStop}%
\bibitem [{\citenamefont {Kilgour}\ and\ \citenamefont
  {Segal}(2018)}]{PhysRevE.98.012117}%
  \BibitemOpen
  \bibfield  {author} {\bibinfo {author} {\bibfnamefont {Michael}\ \bibnamefont
  {Kilgour}}\ and\ \bibinfo {author} {\bibfnamefont {Dvira}\ \bibnamefont
  {Segal}},\ }\bibfield  {title} {\enquote {\bibinfo {title} {Coherence and
  decoherence in quantum absorption refrigerators},}\ }\href {\doibase
  10.1103/PhysRevE.98.012117} {\bibfield  {journal} {\bibinfo  {journal} {Phys.
  Rev. E}\ }\textbf {\bibinfo {volume} {98}},\ \bibinfo {pages} {012117}
  (\bibinfo {year} {2018})}\BibitemShut {NoStop}%
\bibitem [{\citenamefont {Seah}\ \emph {et~al.}(2018)\citenamefont {Seah},
  \citenamefont {Nimmrichter},\ and\ \citenamefont
  {Scarani}}]{PhysRevE.98.012131}%
  \BibitemOpen
  \bibfield  {author} {\bibinfo {author} {\bibfnamefont {Stella}\ \bibnamefont
  {Seah}}, \bibinfo {author} {\bibfnamefont {Stefan}\ \bibnamefont
  {Nimmrichter}}, \ and\ \bibinfo {author} {\bibfnamefont {Valerio}\
  \bibnamefont {Scarani}},\ }\bibfield  {title} {\enquote {\bibinfo {title}
  {Refrigeration beyond weak internal coupling},}\ }\href {\doibase
  10.1103/PhysRevE.98.012131} {\bibfield  {journal} {\bibinfo  {journal} {Phys.
  Rev. E}\ }\textbf {\bibinfo {volume} {98}},\ \bibinfo {pages} {012131}
  (\bibinfo {year} {2018})}\BibitemShut {NoStop}%
\bibitem [{\citenamefont {Mitchison}\ \emph {et~al.}(2016)\citenamefont
  {Mitchison}, \citenamefont {Huber}, \citenamefont {Prior}, \citenamefont
  {Woods},\ and\ \citenamefont {Plenio}}]{Mitchison_2016}%
  \BibitemOpen
  \bibfield  {author} {\bibinfo {author} {\bibfnamefont {Mark~T}\ \bibnamefont
  {Mitchison}}, \bibinfo {author} {\bibfnamefont {Marcus}\ \bibnamefont
  {Huber}}, \bibinfo {author} {\bibfnamefont {Javier}\ \bibnamefont {Prior}},
  \bibinfo {author} {\bibfnamefont {Mischa~P}\ \bibnamefont {Woods}}, \ and\
  \bibinfo {author} {\bibfnamefont {Martin~B}\ \bibnamefont {Plenio}},\
  }\bibfield  {title} {\enquote {\bibinfo {title} {Realising a quantum
  absorption refrigerator with an atom-cavity system},}\ }\href {\doibase
  10.1088/2058-9565/1/1/015001} {\bibfield  {journal} {\bibinfo  {journal}
  {Quant. Sci. Technol.}\ }\textbf {\bibinfo {volume} {1}},\ \bibinfo {pages}
  {015001} (\bibinfo {year} {2016})}\BibitemShut {NoStop}%
\bibitem [{\citenamefont {Chen}\ and\ \citenamefont {Li}(2012)}]{Chen_2012}%
  \BibitemOpen
  \bibfield  {author} {\bibinfo {author} {\bibfnamefont {Yi-Xin}\ \bibnamefont
  {Chen}}\ and\ \bibinfo {author} {\bibfnamefont {Sheng-Wen}\ \bibnamefont
  {Li}},\ }\bibfield  {title} {\enquote {\bibinfo {title} {Quantum refrigerator
  driven by current noise},}\ }\href {\doibase 10.1209/0295-5075/97/40003}
  {\bibfield  {journal} {\bibinfo  {journal} {Europhys. Lett.}\ }\textbf
  {\bibinfo {volume} {97}},\ \bibinfo {pages} {40003} (\bibinfo {year}
  {2012})}\BibitemShut {NoStop}%
\bibitem [{\citenamefont {He}\ \emph {et~al.}(2017)\citenamefont {He},
  \citenamefont {Huang},\ and\ \citenamefont {Yu}}]{PhysRevE.96.052126}%
  \BibitemOpen
  \bibfield  {author} {\bibinfo {author} {\bibfnamefont {Zi-chen}\ \bibnamefont
  {He}}, \bibinfo {author} {\bibfnamefont {Xin-yun}\ \bibnamefont {Huang}}, \
  and\ \bibinfo {author} {\bibfnamefont {Chang-shui}\ \bibnamefont {Yu}},\
  }\bibfield  {title} {\enquote {\bibinfo {title} {Enabling the self-contained
  refrigerator to work beyond its limits by filtering the reservoirs},}\ }\href
  {\doibase 10.1103/PhysRevE.96.052126} {\bibfield  {journal} {\bibinfo
  {journal} {Phys. Rev. E}\ }\textbf {\bibinfo {volume} {96}},\ \bibinfo
  {pages} {052126} (\bibinfo {year} {2017})}\BibitemShut {NoStop}%
\bibitem [{\citenamefont {Mukhopadhyay}\ \emph {et~al.}(2018)\citenamefont
  {Mukhopadhyay}, \citenamefont {Misra}, \citenamefont {Bhattacharya},\ and\
  \citenamefont {Pati}}]{PhysRevE.97.062116}%
  \BibitemOpen
  \bibfield  {author} {\bibinfo {author} {\bibfnamefont {Chiranjib}\
  \bibnamefont {Mukhopadhyay}}, \bibinfo {author} {\bibfnamefont {Avijit}\
  \bibnamefont {Misra}}, \bibinfo {author} {\bibfnamefont {Samyadeb}\
  \bibnamefont {Bhattacharya}}, \ and\ \bibinfo {author} {\bibfnamefont
  {Arun~Kumar}\ \bibnamefont {Pati}},\ }\bibfield  {title} {\enquote {\bibinfo
  {title} {Quantum speed limit constraints on a nanoscale autonomous
  refrigerator},}\ }\href {\doibase 10.1103/PhysRevE.97.062116} {\bibfield
  {journal} {\bibinfo  {journal} {Phys. Rev. E}\ }\textbf {\bibinfo {volume}
  {97}},\ \bibinfo {pages} {062116} (\bibinfo {year} {2018})}\BibitemShut
  {NoStop}%
\bibitem [{\citenamefont {Das}\ \emph {et~al.}(2019)\citenamefont {Das},
  \citenamefont {Misra}, \citenamefont {Pal}, \citenamefont {Sen(De)},\ and\
  \citenamefont {Sen}}]{Das_2019}%
  \BibitemOpen
  \bibfield  {author} {\bibinfo {author} {\bibfnamefont {Sreetama}\
  \bibnamefont {Das}}, \bibinfo {author} {\bibfnamefont {Avijit}\ \bibnamefont
  {Misra}}, \bibinfo {author} {\bibfnamefont {Amit~Kumar}\ \bibnamefont {Pal}},
  \bibinfo {author} {\bibfnamefont {Aditi}\ \bibnamefont {Sen(De)}}, \ and\
  \bibinfo {author} {\bibfnamefont {Ujjwal}\ \bibnamefont {Sen}},\ }\bibfield
  {title} {\enquote {\bibinfo {title} {Necessarily transient quantum
  refrigerator},}\ }\href {\doibase 10.1209/0295-5075/125/20007} {\bibfield
  {journal} {\bibinfo  {journal} {{EPL}}\ }\textbf {\bibinfo {volume} {125}},\
  \bibinfo {pages} {20007} (\bibinfo {year} {2019})}\BibitemShut {NoStop}%
\bibitem [{\citenamefont {Latune}\ \emph {et~al.}(2019)\citenamefont {Latune},
  \citenamefont {Sinayskiy},\ and\ \citenamefont {Petruccione}}]{Latune2019}%
  \BibitemOpen
  \bibfield  {author} {\bibinfo {author} {\bibfnamefont {C.~L.}\ \bibnamefont
  {Latune}}, \bibinfo {author} {\bibfnamefont {I.}~\bibnamefont {Sinayskiy}}, \
  and\ \bibinfo {author} {\bibfnamefont {F.}~\bibnamefont {Petruccione}},\
  }\bibfield  {title} {\enquote {\bibinfo {title} {Quantum coherence, many-body
  correlations, and non-thermal effects for autonomous thermal machines},}\
  }\href {\doibase 10.1038/s41598-019-39300-4} {\bibfield  {journal} {\bibinfo
  {journal} {Sci. Rep.}\ }\textbf {\bibinfo {volume} {9}},\ \bibinfo {pages}
  {3191} (\bibinfo {year} {2019})}\BibitemShut {NoStop}%
\bibitem [{\citenamefont {Hewgill}\ \emph {et~al.}(2020)\citenamefont
  {Hewgill}, \citenamefont {Gonz\'alez}, \citenamefont {Palao}, \citenamefont
  {Alonso}, \citenamefont {Ferraro},\ and\ \citenamefont
  {De~Chiara}}]{PhysRevE.101.012109}%
  \BibitemOpen
  \bibfield  {author} {\bibinfo {author} {\bibfnamefont {Adam}\ \bibnamefont
  {Hewgill}}, \bibinfo {author} {\bibfnamefont {J.~Onam}\ \bibnamefont
  {Gonz\'alez}}, \bibinfo {author} {\bibfnamefont {Jos\'e~P.}\ \bibnamefont
  {Palao}}, \bibinfo {author} {\bibfnamefont {Daniel}\ \bibnamefont {Alonso}},
  \bibinfo {author} {\bibfnamefont {Alessandro}\ \bibnamefont {Ferraro}}, \
  and\ \bibinfo {author} {\bibfnamefont {Gabriele}\ \bibnamefont {De~Chiara}},\
  }\bibfield  {title} {\enquote {\bibinfo {title} {Three-qubit refrigerator
  with two-body interactions},}\ }\href {\doibase 10.1103/PhysRevE.101.012109}
  {\bibfield  {journal} {\bibinfo  {journal} {Phys. Rev. E}\ }\textbf {\bibinfo
  {volume} {101}},\ \bibinfo {pages} {012109} (\bibinfo {year}
  {2020})}\BibitemShut {NoStop}%
\bibitem [{\citenamefont {Maslennikov}\ \emph {et~al.}(2019)\citenamefont
  {Maslennikov}, \citenamefont {Ding}, \citenamefont {Habl{\"u}tzel},
  \citenamefont {Gan}, \citenamefont {Roulet}, \citenamefont {Nimmrichter},
  \citenamefont {Dai}, \citenamefont {Scarani},\ and\ \citenamefont
  {Matsukevich}}]{Maslennikov2019}%
  \BibitemOpen
  \bibfield  {author} {\bibinfo {author} {\bibfnamefont {Gleb}\ \bibnamefont
  {Maslennikov}}, \bibinfo {author} {\bibfnamefont {Shiqian}\ \bibnamefont
  {Ding}}, \bibinfo {author} {\bibfnamefont {Roland}\ \bibnamefont
  {Habl{\"u}tzel}}, \bibinfo {author} {\bibfnamefont {Jaren}\ \bibnamefont
  {Gan}}, \bibinfo {author} {\bibfnamefont {Alexandre}\ \bibnamefont {Roulet}},
  \bibinfo {author} {\bibfnamefont {Stefan}\ \bibnamefont {Nimmrichter}},
  \bibinfo {author} {\bibfnamefont {Jibo}\ \bibnamefont {Dai}}, \bibinfo
  {author} {\bibfnamefont {Valerio}\ \bibnamefont {Scarani}}, \ and\ \bibinfo
  {author} {\bibfnamefont {Dzmitry}\ \bibnamefont {Matsukevich}},\ }\bibfield
  {title} {\enquote {\bibinfo {title} {Quantum absorption refrigerator with
  trapped ions},}\ }\href {\doibase 10.1038/s41467-018-08090-0} {\bibfield
  {journal} {\bibinfo  {journal} {Nat. Commun.}\ }\textbf {\bibinfo {volume}
  {10}},\ \bibinfo {pages} {202} (\bibinfo {year} {2019})}\BibitemShut
  {NoStop}%
\bibitem [{\citenamefont {Correa}\ \emph {et~al.}(2014)\citenamefont {Correa},
  \citenamefont {Palao}, \citenamefont {Alonso},\ and\ \citenamefont
  {Adesso}}]{Correa2014}%
  \BibitemOpen
  \bibfield  {author} {\bibinfo {author} {\bibfnamefont {Luis~A.}\ \bibnamefont
  {Correa}}, \bibinfo {author} {\bibfnamefont {Jos{\'e}~P.}\ \bibnamefont
  {Palao}}, \bibinfo {author} {\bibfnamefont {Daniel}\ \bibnamefont {Alonso}},
  \ and\ \bibinfo {author} {\bibfnamefont {Gerardo}\ \bibnamefont {Adesso}},\
  }\bibfield  {title} {\enquote {\bibinfo {title} {Quantum-enhanced absorption
  refrigerators},}\ }\href {\doibase 10.1038/srep03949} {\bibfield  {journal}
  {\bibinfo  {journal} {Sci. Rep.}\ }\textbf {\bibinfo {volume} {4}},\ \bibinfo
  {pages} {3949} (\bibinfo {year} {2014})}\BibitemShut {NoStop}%
\bibitem [{\citenamefont {Fr\"{o}hlich}(1952)}]{frochlich_interaction_1952}%
  \BibitemOpen
  \bibfield  {author} {\bibinfo {author} {\bibfnamefont {H}~\bibnamefont
  {Fr\"{o}hlich}},\ }\bibfield  {title} {\enquote {\bibinfo {title}
  {Interaction of electrons with lattice vibrations},}\ }\href
  {https://royalsocietypublishing.org/doi/abs/10.1098/rspa.1952.0212}
  {\bibfield  {journal} {\bibinfo  {journal} {Proceedings of the Royal Society
  of London. Series A. Mathematical and Physical Sciences}\ } (\bibinfo {year}
  {1952})}\BibitemShut {NoStop}%
\bibitem [{\citenamefont {Holstein}(1959)}]{holstein_studiesI_1959}%
  \BibitemOpen
  \bibfield  {author} {\bibinfo {author} {\bibfnamefont {T}~\bibnamefont
  {Holstein}},\ }\bibfield  {title} {\enquote {\bibinfo {title} {Studies of
  polaron motion: {Part} {I}. {The} molecular-crystal model},}\ }\href
  {\doibase 10.1016/0003-4916(59)90002-8} {\bibfield  {journal} {\bibinfo
  {journal} {Annals of Physics}\ }\textbf {\bibinfo {volume} {8}},\ \bibinfo
  {pages} {325--342} (\bibinfo {year} {1959})}\BibitemShut {NoStop}%
\bibitem [{\citenamefont {Weiss}(2012)}]{weiss_quantumDissSys_4ed2012}%
  \BibitemOpen
  \bibfield  {author} {\bibinfo {author} {\bibfnamefont {U}~\bibnamefont
  {Weiss}},\ }\href
  {https://www.worldscientific.com/worldscibooks/10.1142/8334} {\emph {\bibinfo
  {title} {Quantum Dissipative Systems}}},\ \bibinfo {edition} {4th}\ ed.\
  (\bibinfo  {publisher} {World Scientific},\ \bibinfo {address} {New Jersey},\
  \bibinfo {year} {2012})\BibitemShut {NoStop}%
\bibitem [{\citenamefont {Brunner}\ \emph {et~al.}(2014)\citenamefont
  {Brunner}, \citenamefont {Huber}, \citenamefont {Linden}, \citenamefont
  {Popescu}, \citenamefont {Silva},\ and\ \citenamefont
  {Skrzypczyk}}]{PhysRevE.89.032115}%
  \BibitemOpen
  \bibfield  {author} {\bibinfo {author} {\bibfnamefont {Nicolas}\ \bibnamefont
  {Brunner}}, \bibinfo {author} {\bibfnamefont {Marcus}\ \bibnamefont {Huber}},
  \bibinfo {author} {\bibfnamefont {Noah}\ \bibnamefont {Linden}}, \bibinfo
  {author} {\bibfnamefont {Sandu}\ \bibnamefont {Popescu}}, \bibinfo {author}
  {\bibfnamefont {Ralph}\ \bibnamefont {Silva}}, \ and\ \bibinfo {author}
  {\bibfnamefont {Paul}\ \bibnamefont {Skrzypczyk}},\ }\bibfield  {title}
  {\enquote {\bibinfo {title} {Entanglement enhances cooling in microscopic
  quantum refrigerators},}\ }\href {\doibase 10.1103/PhysRevE.89.032115}
  {\bibfield  {journal} {\bibinfo  {journal} {Phys. Rev. E}\ }\textbf {\bibinfo
  {volume} {89}},\ \bibinfo {pages} {032115} (\bibinfo {year}
  {2014})}\BibitemShut {NoStop}%
\bibitem [{\citenamefont {Karg{\i}}\ \emph {et~al.}(2019)\citenamefont
  {Karg{\i}}, \citenamefont {Naseem}, \citenamefont {Opatrn\'{y}},\ and\
  \citenamefont {Kurizki}}]{PhysRevE.99.042121}%
  \BibitemOpen
  \bibfield  {author} {\bibinfo {author} {\bibfnamefont {Cahit}\ \bibnamefont
  {Karg{\i}}}, \bibinfo {author} {\bibfnamefont {M.~Tahir}\ \bibnamefont
  {Naseem}}, \bibinfo {author} {\bibfnamefont {\"{O}zg\"{u}r~E.}\ \bibnamefont
  {Opatrn\'{y}}, \bibfnamefont {Tom\'{a}\v{s}~andM\"{u}stecapl{\i}\u{g}lu}}, \
  and\ \bibinfo {author} {\bibfnamefont {Gershon}\ \bibnamefont {Kurizki}},\
  }\bibfield  {title} {\enquote {\bibinfo {title} {Quantum optical two-atom
  thermal diode},}\ }\href {\doibase 10.1103/PhysRevE.99.042121} {\bibfield
  {journal} {\bibinfo  {journal} {Phys. Rev. E}\ }\textbf {\bibinfo {volume}
  {99}},\ \bibinfo {pages} {042121} (\bibinfo {year} {2019})}\BibitemShut
  {NoStop}%
\bibitem [{\citenamefont {Naseem}\ \emph {et~al.}(2018)\citenamefont {Naseem},
  \citenamefont {Xuereb},\ and\ \citenamefont
  {M\"{u}stecapl{\i}\u{g}lu}}]{PhysRevA.98.052123}%
  \BibitemOpen
  \bibfield  {author} {\bibinfo {author} {\bibfnamefont {M.~Tahir}\
  \bibnamefont {Naseem}}, \bibinfo {author} {\bibfnamefont {Andr\'{e}}\
  \bibnamefont {Xuereb}}, \ and\ \bibinfo {author} {\bibfnamefont
  {\"{O}zg\"{u}r~E.}\ \bibnamefont {M\"{u}stecapl{\i}\u{g}lu}},\ }\bibfield
  {title} {\enquote {\bibinfo {title} {Thermodynamic consistency of the
  optomechanical master equation},}\ }\href {\doibase
  10.1103/PhysRevA.98.052123} {\bibfield  {journal} {\bibinfo  {journal} {Phys.
  Rev. A}\ }\textbf {\bibinfo {volume} {98}},\ \bibinfo {pages} {052123}
  (\bibinfo {year} {2018})}\BibitemShut {NoStop}%
\bibitem [{\citenamefont {Aspelmeyer}\ \emph {et~al.}(2014)\citenamefont
  {Aspelmeyer}, \citenamefont {Kippenberg},\ and\ \citenamefont
  {Marquardt}}]{RevModPhys.86.1391}%
  \BibitemOpen
  \bibfield  {author} {\bibinfo {author} {\bibfnamefont {Markus}\ \bibnamefont
  {Aspelmeyer}}, \bibinfo {author} {\bibfnamefont {Tobias~J.}\ \bibnamefont
  {Kippenberg}}, \ and\ \bibinfo {author} {\bibfnamefont {Florian}\
  \bibnamefont {Marquardt}},\ }\bibfield  {title} {\enquote {\bibinfo {title}
  {Cavity optomechanics},}\ }\href {\doibase 10.1103/RevModPhys.86.1391}
  {\bibfield  {journal} {\bibinfo  {journal} {Rev. Mod. Phys.}\ }\textbf
  {\bibinfo {volume} {86}},\ \bibinfo {pages} {1391--1452} (\bibinfo {year}
  {2014})}\BibitemShut {NoStop}%
\bibitem [{\citenamefont {Gelbwaser-Klimovsky}\ \emph
  {et~al.}(2013)\citenamefont {Gelbwaser-Klimovsky}, \citenamefont {Alicki},\
  and\ \citenamefont {Kurizki}}]{PhysRevE.87.012140}%
  \BibitemOpen
  \bibfield  {author} {\bibinfo {author} {\bibfnamefont {D.}~\bibnamefont
  {Gelbwaser-Klimovsky}}, \bibinfo {author} {\bibfnamefont {R.}~\bibnamefont
  {Alicki}}, \ and\ \bibinfo {author} {\bibfnamefont {G.}~\bibnamefont
  {Kurizki}},\ }\bibfield  {title} {\enquote {\bibinfo {title} {Minimal
  universal quantum heat machine},}\ }\href {\doibase
  10.1103/PhysRevE.87.012140} {\bibfield  {journal} {\bibinfo  {journal} {Phys.
  Rev. E}\ }\textbf {\bibinfo {volume} {87}},\ \bibinfo {pages} {012140}
  (\bibinfo {year} {2013})}\BibitemShut {NoStop}%
\bibitem [{\citenamefont {Leggett}\ \emph {et~al.}(1987)\citenamefont
  {Leggett}, \citenamefont {Chakravarty}, \citenamefont {Dorsey}, \citenamefont
  {Fisher}, \citenamefont {Garg},\ and\ \citenamefont
  {Zwerger}}]{RevModPhys.59.1}%
  \BibitemOpen
  \bibfield  {author} {\bibinfo {author} {\bibfnamefont {A.~J.}\ \bibnamefont
  {Leggett}}, \bibinfo {author} {\bibfnamefont {S.}~\bibnamefont
  {Chakravarty}}, \bibinfo {author} {\bibfnamefont {A.~T.}\ \bibnamefont
  {Dorsey}}, \bibinfo {author} {\bibfnamefont {Matthew P.~A.}\ \bibnamefont
  {Fisher}}, \bibinfo {author} {\bibfnamefont {Anupam}\ \bibnamefont {Garg}}, \
  and\ \bibinfo {author} {\bibfnamefont {W.}~\bibnamefont {Zwerger}},\
  }\bibfield  {title} {\enquote {\bibinfo {title} {Dynamics of the dissipative
  two-state system},}\ }\href {\doibase 10.1103/RevModPhys.59.1} {\bibfield
  {journal} {\bibinfo  {journal} {Rev. Mod. Phys.}\ }\textbf {\bibinfo {volume}
  {59}},\ \bibinfo {pages} {1--85} (\bibinfo {year} {1987})}\BibitemShut
  {NoStop}%
\bibitem [{\citenamefont {Kofman}\ \emph {et~al.}(1994)\citenamefont {Kofman},
  \citenamefont {Kurizki},\ and\ \citenamefont
  {Sherman}}]{doi:10.1080/09500349414550381}%
  \BibitemOpen
  \bibfield  {author} {\bibinfo {author} {\bibfnamefont {A.G.}\ \bibnamefont
  {Kofman}}, \bibinfo {author} {\bibfnamefont {G.}~\bibnamefont {Kurizki}}, \
  and\ \bibinfo {author} {\bibfnamefont {B.}~\bibnamefont {Sherman}},\
  }\bibfield  {title} {\enquote {\bibinfo {title} {Spontaneous and induced
  atomic decay in photonic band structures},}\ }\href {\doibase
  10.1080/09500349414550381} {\bibfield  {journal} {\bibinfo  {journal} {J.
  Mod. Opt.}\ }\textbf {\bibinfo {volume} {41}},\ \bibinfo {pages} {353--384}
  (\bibinfo {year} {1994})}\BibitemShut {NoStop}%
\bibitem [{\citenamefont {Gelbwaser-Klimovsky}\ and\ \citenamefont
  {Kurizki}(2014)}]{PhysRevE.90.022102}%
  \BibitemOpen
  \bibfield  {author} {\bibinfo {author} {\bibfnamefont {D.}~\bibnamefont
  {Gelbwaser-Klimovsky}}\ and\ \bibinfo {author} {\bibfnamefont
  {G.}~\bibnamefont {Kurizki}},\ }\bibfield  {title} {\enquote {\bibinfo
  {title} {Heat-machine control by quantum-state preparation: From quantum
  engines to refrigerators},}\ }\href {\doibase 10.1103/PhysRevE.90.022102}
  {\bibfield  {journal} {\bibinfo  {journal} {Phys. Rev. E}\ }\textbf {\bibinfo
  {volume} {90}},\ \bibinfo {pages} {022102} (\bibinfo {year}
  {2014})}\BibitemShut {NoStop}%
\bibitem [{\citenamefont {Naseem}\ \emph {et~al.}(2020)\citenamefont {Naseem},
  \citenamefont {Misra}, \citenamefont {\"{O}zg\"{u}r
  E.~M\"{u}stecapl{\i}\u{g}lu},\ and\ \citenamefont {Kurizki}}]{MF}%
  \BibitemOpen
  \bibfield  {author} {\bibinfo {author} {\bibfnamefont {M.~Tahir}\
  \bibnamefont {Naseem}}, \bibinfo {author} {\bibfnamefont {Avijit}\
  \bibnamefont {Misra}}, \bibinfo {author} {\bibnamefont {\"{O}zg\"{u}r
  E.~M\"{u}stecapl{\i}\u{g}lu}}, \ and\ \bibinfo {author} {\bibfnamefont
  {Gershon}\ \bibnamefont {Kurizki}},\ }\bibfield  {title} {\enquote {\bibinfo
  {title} {Minimal quantum heat manager boosted by bath spectral filtering},}\
  }\href {https://arxiv.org/abs/2004.07393} {\bibfield  {journal} {\bibinfo
  {journal} {arXiv:2004.07393}\ } (\bibinfo {year} {2020})}\BibitemShut
  {NoStop}%
\bibitem [{\citenamefont {Nieuwenhuizen}\ and\ \citenamefont
  {Allahverdyan}(2002)}]{PhysRevE.66.036102}%
  \BibitemOpen
  \bibfield  {author} {\bibinfo {author} {\bibfnamefont {Th.~M.}\ \bibnamefont
  {Nieuwenhuizen}}\ and\ \bibinfo {author} {\bibfnamefont {A.~E.}\ \bibnamefont
  {Allahverdyan}},\ }\bibfield  {title} {\enquote {\bibinfo {title}
  {Statistical thermodynamics of quantum brownian motion: Construction of
  perpetuum mobile of the second kind},}\ }\href {\doibase
  10.1103/PhysRevE.66.036102} {\bibfield  {journal} {\bibinfo  {journal} {Phys.
  Rev. E}\ }\textbf {\bibinfo {volume} {66}},\ \bibinfo {pages} {036102}
  (\bibinfo {year} {2002})}\BibitemShut {NoStop}%
\bibitem [{\citenamefont {Geusic}\ \emph {et~al.}(1967)\citenamefont {Geusic},
  \citenamefont {Schulz-DuBios},\ and\ \citenamefont
  {Scovil}}]{PhysRev.156.343}%
  \BibitemOpen
  \bibfield  {author} {\bibinfo {author} {\bibfnamefont {J.~E.}\ \bibnamefont
  {Geusic}}, \bibinfo {author} {\bibfnamefont {E.~O.}\ \bibnamefont
  {Schulz-DuBios}}, \ and\ \bibinfo {author} {\bibfnamefont {H.~E.~D.}\
  \bibnamefont {Scovil}},\ }\bibfield  {title} {\enquote {\bibinfo {title}
  {Quantum equivalent of the carnot cycle},}\ }\href {\doibase
  10.1103/PhysRev.156.343} {\bibfield  {journal} {\bibinfo  {journal} {Phys.
  Rev.}\ }\textbf {\bibinfo {volume} {156}},\ \bibinfo {pages} {343--351}
  (\bibinfo {year} {1967})}\BibitemShut {NoStop}%
\bibitem [{\citenamefont {Peres}(1996)}]{PhysRevLett.77.1413}%
  \BibitemOpen
  \bibfield  {author} {\bibinfo {author} {\bibfnamefont {Asher}\ \bibnamefont
  {Peres}},\ }\bibfield  {title} {\enquote {\bibinfo {title} {Separability
  criterion for density matrices},}\ }\href {\doibase
  10.1103/PhysRevLett.77.1413} {\bibfield  {journal} {\bibinfo  {journal}
  {Phys. Rev. Lett.}\ }\textbf {\bibinfo {volume} {77}},\ \bibinfo {pages}
  {1413--1415} (\bibinfo {year} {1996})}\BibitemShut {NoStop}%
\bibitem [{\citenamefont {Ollivier}\ and\ \citenamefont
  {Zurek}(2001)}]{PhysRevLett.88.017901}%
  \BibitemOpen
  \bibfield  {author} {\bibinfo {author} {\bibfnamefont {Harold}\ \bibnamefont
  {Ollivier}}\ and\ \bibinfo {author} {\bibfnamefont {Wojciech~H.}\
  \bibnamefont {Zurek}},\ }\bibfield  {title} {\enquote {\bibinfo {title}
  {Quantum discord: A measure of the quantumness of correlations},}\ }\href
  {\doibase 10.1103/PhysRevLett.88.017901} {\bibfield  {journal} {\bibinfo
  {journal} {Phys. Rev. Lett.}\ }\textbf {\bibinfo {volume} {88}},\ \bibinfo
  {pages} {017901} (\bibinfo {year} {2001})}\BibitemShut {NoStop}%
\bibitem [{\citenamefont {Henderson}\ and\ \citenamefont
  {Vedral}(2001)}]{Henderson_2001}%
  \BibitemOpen
  \bibfield  {author} {\bibinfo {author} {\bibfnamefont {L}~\bibnamefont
  {Henderson}}\ and\ \bibinfo {author} {\bibfnamefont {V}~\bibnamefont
  {Vedral}},\ }\bibfield  {title} {\enquote {\bibinfo {title} {Classical,
  quantum and total correlations},}\ }\href {\doibase
  10.1088/0305-4470/34/35/315} {\bibfield  {journal} {\bibinfo  {journal}
  {Journal of Physics A: Mathematical and General}\ }\textbf {\bibinfo {volume}
  {34}},\ \bibinfo {pages} {6899--6905} (\bibinfo {year} {2001})}\BibitemShut
  {NoStop}%
\bibitem [{\citenamefont {Huang}(2014)}]{Huang_2014}%
  \BibitemOpen
  \bibfield  {author} {\bibinfo {author} {\bibfnamefont {Yichen}\ \bibnamefont
  {Huang}},\ }\bibfield  {title} {\enquote {\bibinfo {title} {Computing quantum
  discord is {NP}-complete},}\ }\href {\doibase 10.1088/1367-2630/16/3/033027}
  {\bibfield  {journal} {\bibinfo  {journal} {New Journal of Physics}\ }\textbf
  {\bibinfo {volume} {16}},\ \bibinfo {pages} {033027} (\bibinfo {year}
  {2014})}\BibitemShut {NoStop}%
\bibitem [{\citenamefont {Giorda}\ and\ \citenamefont
  {Paris}(2010)}]{PhysRevLett.105.020503}%
  \BibitemOpen
  \bibfield  {author} {\bibinfo {author} {\bibfnamefont {Paolo}\ \bibnamefont
  {Giorda}}\ and\ \bibinfo {author} {\bibfnamefont {Matteo G.~A.}\ \bibnamefont
  {Paris}},\ }\bibfield  {title} {\enquote {\bibinfo {title} {Gaussian quantum
  discord},}\ }\href {\doibase 10.1103/PhysRevLett.105.020503} {\bibfield
  {journal} {\bibinfo  {journal} {Phys. Rev. Lett.}\ }\textbf {\bibinfo
  {volume} {105}},\ \bibinfo {pages} {020503} (\bibinfo {year}
  {2010})}\BibitemShut {NoStop}%
\bibitem [{\citenamefont {Correa}\ \emph {et~al.}(2015)\citenamefont {Correa},
  \citenamefont {Palao},\ and\ \citenamefont {Alonso}}]{PhysRevE.92.032136}%
  \BibitemOpen
  \bibfield  {author} {\bibinfo {author} {\bibfnamefont {Luis~A.}\ \bibnamefont
  {Correa}}, \bibinfo {author} {\bibfnamefont {Jos\'e~P.}\ \bibnamefont
  {Palao}}, \ and\ \bibinfo {author} {\bibfnamefont {Daniel}\ \bibnamefont
  {Alonso}},\ }\bibfield  {title} {\enquote {\bibinfo {title} {Internal
  dissipation and heat leaks in quantum thermodynamic cycles},}\ }\href
  {\doibase 10.1103/PhysRevE.92.032136} {\bibfield  {journal} {\bibinfo
  {journal} {Phys. Rev. E}\ }\textbf {\bibinfo {volume} {92}},\ \bibinfo
  {pages} {032136} (\bibinfo {year} {2015})}\BibitemShut {NoStop}%
\bibitem [{\citenamefont {Eichler}\ and\ \citenamefont
  {Petta}(2018)}]{PhysRevLett.120.227702}%
  \BibitemOpen
  \bibfield  {author} {\bibinfo {author} {\bibfnamefont {C.}~\bibnamefont
  {Eichler}}\ and\ \bibinfo {author} {\bibfnamefont {J.~R.}\ \bibnamefont
  {Petta}},\ }\bibfield  {title} {\enquote {\bibinfo {title} {Realizing a
  circuit analog of an optomechanical system with longitudinally coupled
  superconducting resonators},}\ }\href {\doibase
  10.1103/PhysRevLett.120.227702} {\bibfield  {journal} {\bibinfo  {journal}
  {Phys. Rev. Lett.}\ }\textbf {\bibinfo {volume} {120}},\ \bibinfo {pages}
  {227702} (\bibinfo {year} {2018})}\BibitemShut {NoStop}%
\bibitem [{\citenamefont {Mari}\ and\ \citenamefont
  {Eisert}(2012)}]{PhysRevLett.108.120602}%
  \BibitemOpen
  \bibfield  {author} {\bibinfo {author} {\bibfnamefont {A.}~\bibnamefont
  {Mari}}\ and\ \bibinfo {author} {\bibfnamefont {J.}~\bibnamefont {Eisert}},\
  }\bibfield  {title} {\enquote {\bibinfo {title} {Cooling by heating: Very hot
  thermal light can significantly cool quantum systems},}\ }\href {\doibase
  10.1103/PhysRevLett.108.120602} {\bibfield  {journal} {\bibinfo  {journal}
  {Phys. Rev. Lett.}\ }\textbf {\bibinfo {volume} {108}},\ \bibinfo {pages}
  {120602} (\bibinfo {year} {2012})}\BibitemShut {NoStop}%
\end{thebibliography}
\end{document}